\documentclass[showpacs,preprintnumbers,amsmath,amssymb,twocolumn,superscriptaddress,prb]{revtex4}
\usepackage{graphicx}
\usepackage{amsfonts}
\usepackage{amsmath, amsthm, amssymb}
\usepackage{dsfont}
\usepackage{times}

\newcommand{\ve}[1]{\boldsymbol{#1}}
\newcommand{\va}{{\ve{a}}} 
\newcommand{\vk}{{\ve{k}}} 
\newcommand{\vq}{{\ve{q}}} 
\newcommand{\DeltaJ}{\ve{\Delta}_J} 

\newcommand{\e}[1]{\mathrm{e}^{#1}}

\newcommand{\swave}{$s$-wave}
\newcommand{\dwave}{$d$-wave}
\newcommand{\daa}{$d_{x^2-y^2}$-wave}
\newcommand{\dbb}{$d_{xy}$-wave}

\newcommand{\etal}{\emph{et al.}}
\def\i{\mathrm{i}}

\begin{document}
\title[Josephson current in graphene: the role of unconventional pairing symmetries]{Josephson current in graphene: the role of unconventional pairing symmetries}
\author{Jacob Linder*}
\affiliation{Department of Physics, Norwegian University of
Science and Technology, N-7491 Trondheim, Norway}
\author{Annica M. Black-Schaffer*}
\affiliation{Department of Applied Physics, Stanford University, Stanford, California 94305, USA}
\author{Takehito Yokoyama}
\affiliation{Department of Applied Physics, University of Tokyo, Tokyo 113-8656, Japan}
\author{Sebastian Doniach}
\affiliation{Departments of Physics and Applied Physics, Stanford University, Stanford, California 94305, USA}
\author{Asle Sudb{\o}}
\affiliation{Department of Physics, Norwegian University of
Science and Technology, N-7491 Trondheim, Norway\\
* These authors contributed equally to this work}

\date{Received \today}
\begin{abstract}
\noindent We investigate the Josephson current in a graphene superconductor/normal/superconductor junction, where superconductivity is induced by means of the proximity effect from external contacts. We take into account the possibility of anisotropic pairing by also including singlet nearest-neighbor interactions, and investigate how the transport properties are affected by the symmetry of the superconducting order parameter. This corresponds to an extension of the usual on-site interaction assumption, which yields an isotropic $s$-wave order parameter near the Dirac points. Here, we employ a full numerical solution as well as an analytical treatment, and show how the proximity effect may induce exotic types of superconducting states near the Dirac points, e.g.~$p_x$- and $p_y$-wave pairing or a combination of $s$-wave and $p+\i p$-wave pairing. We find that the Josephson current exhibits a weakly-damped, oscillatory dependence on the length of the junction when the graphene sheet is strongly doped. The analytical and numerical treatments are found to agree well with each other in the $s$-wave case when calculating the critical current and current-phase relationship. For the scenarios with anisotropic superconducting pairing, there is a deviation between the two treatments, especially for the effective $p_x$-wave order parameter near the Dirac cones which features zero-energy states at the interfaces. This indicates that a numerical, self-consistent approach becomes necessary when treating anisotropic superconducting pairing in graphene. 
 \end{abstract}

\maketitle
\section{Introduction}
The unusual electronic properties of the charge-carriers in graphene\cite{novoselov_science_04} have triggered a massive interest in this material over the last few years. Graphene is a monolayer of graphite, and thus has a two-dimensional honeycomb lattice structure consisting of two triangular sublattices. The two most interesting features of the dispersion relation for the quasiparticles moving in a graphene sheet is that \textit{(i)} the quasiparticles at Fermi level are nodal, meaning there is no Fermi surface at zero doping, and that \textit{(ii)} the band structure is conical, thus giving rise to an effective mass of zero for the quasiparticles. These facts have paramount implications for a number of physical properties of a graphene sheet.\cite{castro_rmp_09, beenakker_rmp_08}
\par
Quite recently, proximity-induced superconductivity in graphene was achieved experimentally by means of depositing superconducting contacts on a graphene sheet.\cite{heersche_nature_06, du_prb_08, ojeda_prb_09} This has led to multiple investigations with respect to the transport properties of superconducting graphene.\cite{beenakker_prl_06, titov_prb_06, bhattacharjee_prl_06, moghaddam_prb_06, linder_prl_07, black-schaffer_prb_08, bhattacharjee_prb_07, maiti_prb_07, linder_prb_08, yokoyama_prb_08,Burset, rainis_prb_09, jiang_prb_08, liang_prl_08} Some of the key findings in these investigations include the possibility of specular Andreev reflection,\cite{beenakker_prl_06} oscillations in the conductance of a superconductor/normal (SN) junction,\cite{bhattacharjee_prl_06, linder_prl_07} and a finite Josephson current even at the Dirac point in an SNS junction.\cite{titov_prb_06} Very recently, some authors have also explored the interplay between proximity-induced ferromagnetism\cite{Haugen,Yokoyama} and superconductivity in graphene.\cite{linder_prl_08, asano_prb_08, zhang_prl_08, moghaddam_prb_08} So far in the literature, conventional $s$-wave superconducting pairing has been the primary focus, whereas only little attention has been paid to how unconventional pairing in superconducting graphene structures influence the transport properties.\cite{jiang_prb_08, linder_prb_08}
\par
In the majority of studies considering transport properties of superconducting graphene hybrid structures only an analytical scattering matrix approach has been employed. The advantages of such a treatment as compared to a purely numerical one is that it often offers more physical insight into the problem. On the other hand, a numerical self-consistent treatment is more authoritative and will in general provide more accurate results both qualitatively and quantitatively.\cite{black-schaffer_prb_08} Ideally, it would therefore be desirable to compare an analytical approach with a numerical treatment in order to see which conclusions obtained in the former case still hold in the latter case. 
\par
\begin{figure}[t!]
\centering
\resizebox{0.4\textwidth}{!}{
\includegraphics{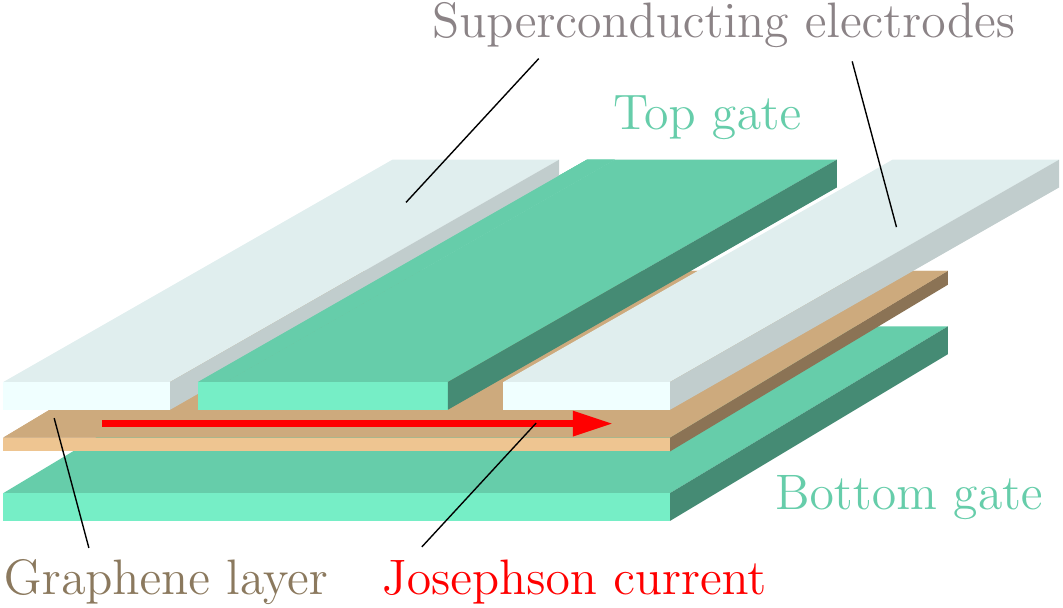}}
\caption{(Color online) The experimental setup proposed in this paper. Two superconducting electrodes in close proximity to a graphene sheet induces superconductivity. The Josephson current flows through the graphene sheet indicated by the red arrow. Top and bottom gates contacted to the graphene sheet permit local control over the chemical potential.}
\label{fig:model}
\end{figure}

Motivated by this, we present in this paper both an analytical and numerical study of the Josephson current in an SNS graphene junction (see Fig. \ref{fig:model}) using both conventional and unconventional superconducting contacts. We here allow for the unconventional pairing by means of including nearest-neighbor interactions in a tight-binding model. This interaction gives rise to exotic types of superconducting states near the Dirac points, which also significantly alters the behavior of the supercurrent in the system. Our main result is that, whereas the analytical and numerical treatments are found to agree well with each other in the isotropic $s$-wave case, there is a deviation between the two treatments in the anisotropic case, especially for an effective $p_x$-wave order parameter near the Dirac cones featuring zero-energy states. This finding suggests that a numerical, self-consistent approach is required when studying anisotropic superconducting pairing in graphene. 
\par
The article is organized as follows. In Sec.~\ref{sec:theory}, we lay the theoretical foundation in terms of notation and formalism. In Sec.~\ref{sec:results}, we present and discuss our main results, focusing on the analytical treatment in Sec.~\ref{sec:results_analytical} and the numerical approach in \ref{sec:results_numerical}. Finally, we summarize our findings in Sec.~\ref{sec:summary}.

\section{Theory}\label{sec:theory}

\subsection{Analytical treatment}
\begin{figure}[h!]
\centering
\includegraphics{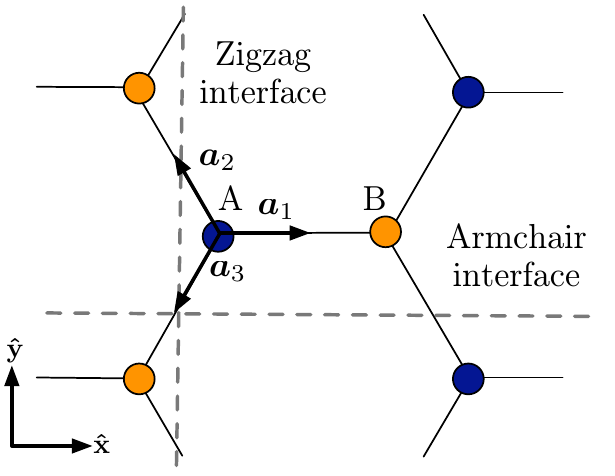}
\caption{(Color online) The graphene honeycomb lattice with the two different atomic sites $A$ and $B$, the three nearest neighbor directions $\{\boldsymbol{a}_1,\boldsymbol{a}_2,\boldsymbol{a}_3\}$, and the zigzag and armchair interfaces marked.
}
\label{fig:graphene}
\end{figure}

Our starting point is the tight-binding Hamiltonian on the graphene lattice, including a superconducting pairing order parameter which is induced by the proximity effect from external superconducting contacts. We include the possibility for both on-site superconducting pairing and nearest-neighbor pairing by means of nearest neighbor spin-singlet bond (SB) correlations. The full Hamiltonian takes the form:
\begin{align}
\label{eq:Hana}
H &= -t\sum_{ij\sigma} (A_{i\sigma}^\dag B_{i+\boldsymbol{a}_j,\sigma} + \text{H.c}) \notag\\
&- \mu\sum_{i\sigma} (A_{\i\sigma}^\dag A_{\i\sigma} + B_{i,\sigma}^\dag B_{i,\sigma}) \notag\\
&+ \sum_{ij} [\Delta_{J\boldsymbol{a}_j}(A_{i\uparrow}^\dag B_{i+\boldsymbol{a}_j,\downarrow}^\dag - A_{i\downarrow}^\dag B_{i+\boldsymbol{a}_j,\uparrow}^\dag)+ \text{H.c.}]\notag\\
&+ \sum_i [\Delta_U (A_{i\uparrow}^\dag A_{i\downarrow}^\dag + B_{i\uparrow}^\dag B_{i\downarrow}^\dag) + \text{H.c.}]\notag\\
&+ \text{constant terms}.
\end{align}
Here, $A_{i\sigma}$ and $B_{i\sigma}$ are the second quantized fermion operators on the sublattices $A$ and $B$, while $\boldsymbol{a}_j$ denotes the three nearest-neighbor vectors, see Fig.~\ref{fig:graphene}.
 They read $\boldsymbol{a}_1 = a(1,0)$, $\boldsymbol{a}_2 = a(-1,\sqrt{3})/2$, $\boldsymbol{a}_3 = a(-1,-\sqrt{3})/2$, where $a$ is 
 the inter-atomic distance of the carbon atoms. For later use, we also define the reciprocal vectors to the Dirac points $K_\pm$ 
 as $\boldsymbol{K}_\pm = (0,\pm \frac{4\pi}{3\sqrt{3}a})$. The superconducting pairing is accounted for by $\Delta_U$, which 
 corresponds to the on-site interaction, and $\Delta_{J\boldsymbol{a}_j}$, which corresponds to the pairing interaction along 
 the nearest-neighbor vectors $\boldsymbol{a}_j$. $\Delta_{J\boldsymbol{a}_j}$ may in general be different for different 
 $\boldsymbol{a}_j$'s, i.e.~the three different nearest neighbor 
 bonds. In fact, a self-consistent solution admits three possible solutions\cite{black-schaffer_prb_07} classified as follows upon 
 defining $\boldsymbol{\Delta}_J = (\Delta_{J\boldsymbol{a}_1},\Delta_{J\boldsymbol{a}_2},\Delta_{J\boldsymbol{a}_3})$:
\begin{align}
\label{eq:sym}
\text{Extended $s$-wave:}\; &\boldsymbol{\Delta}_J = \Delta_t(1,1,1),\notag\\
\text{$d_{x^2-y^2}$-wave:}\; &\boldsymbol{\Delta}_J = \Delta_t(2,-1,-1),\notag\\
\text{$d_{xy}$-wave:}\; &\boldsymbol{\Delta}_J = \Delta_t(0,1,-1).
\end{align}
The classification of the different symmetries stems from which irreducible representation of the crystal point group $D_{6}$ 
they belong to when considered in the whole band structure Brillouin zone, i.e.~the full reciprocal unit cell in the basis 
where the kinetic energy is diagonal. The extended $s$-wave gap is proportional to the band dispersion, i.e.~ $\propto \epsilon_\vq$ 
given below in Eq.~(\ref{eq:bandstr}). It has the full symmetry of the lattice, thus belonging to the A$_1$ irreducible representation 
although, in contrast to the on-site $s$-wave $\Delta_U$, it varies in magnitude over the Fermi surface. The different $d$-wave solutions 
have four-fold symmetries and belong to the two-dimensional E$_2$ irreducible representation and, therefore, technically any linear 
combination of these two solutions is a valid solution from a symmetry standpoint. For an effective potential giving rise to an 
intrinsic SB pairing in the translational invariant bulk the two $d$-wave states are degenerate at $T_c$ but the complex combination 
$d_{x^2-y^2 } + \i d_{xy}$ is favored just below $T_c$.\cite{black-schaffer_prb_07} Interestingly, this state breaks thus time-reversal 
symmetry (TRS). Here, however, the SB pairing is induced into the graphene from external contacts and we choose to limit the 
symmetries studied to the ones given in Eq.~(\ref{eq:sym}) as those would be the ones naturally induced from correspondingly 
aligned $d$-wave superconducting contacts, such as high-$T_c$ cuprate superconductors. 
\par
Introducing the Fourier transform of the fermion operators according to
\begin{align}
A_{\vq\sigma} = \sum_i A_{i\sigma} \e{-\i\vq\boldsymbol{r}_i},\; A_{i\sigma} = \sum_\vq A_{\vq\sigma} \e{\i\vq\boldsymbol{r}_i},
\end{align}
and similarly for $A\to B$, we may write down the Hamiltonian in momentum space (now discarding irrelevant constant terms):
\begin{align}\label{eq:H}
H &= \sum_\vq \varphi_\vq^\dag M_\vk \varphi_\vq,\notag\\
\varphi_\vq^\dag &= (A_{\vq\uparrow}^\dag, B_{\vq\uparrow}^\dag, A_{-\vq\downarrow}, B_{-\vq\downarrow}),\notag\\
M_\vq &= \begin{pmatrix}
-\mu & \varepsilon(\vq) & \Delta_U & \Delta(\vq) \\
\varepsilon^\dag(\vq) & -\mu & \Delta(-\vq) & \Delta_U \\
\Delta_U^\dag & \Delta^\dag(-\vq) & \mu & -\varepsilon^\dag(-\vq) \\
\Delta^\dag(\vq) & \Delta_U^\dag & -\varepsilon(-\vq) & \mu \\
\end{pmatrix}.
\end{align}
Again, $\Delta_U$ is the superconducting order parameter resulting from on-site pairing interaction, while we have 
defined
\begin{align}
\varepsilon(\vq) = -t\sum_j \e{\i\vq\boldsymbol{a}_j},\; \Delta(\vq) = \sum_j \Delta_{J\boldsymbol{a}_j} \e{\i\vq\boldsymbol{a}_j}.
\end{align}
We are interested in the behaviour near the Dirac points $K_\pm$, and hence wish to evaluate $M_\vq$ at 
$\vq=\boldsymbol{K}_\pm + \vk$ where $|\vk| \ll |\boldsymbol{K}_\pm|$, i.e.~in the low-energy limit. 
In this case, we find by means of a straight-forward Taylor expansion, that 
\begin{align}
\varepsilon(\boldsymbol{K}_\pm + \vk) = v_F(-\i k_x \pm k_y),\notag\\
\end{align}
with the definition $v_F=3ta/2$, while for the superconducting order parameters we have
\begin{align}\label{eq:syms}
\text{Extended $s$-wave:}\; &\Delta(\boldsymbol{K}_\pm + \vk) = \frac{3\Delta_ta}{2}(\i k_x \mp k_y),\notag\\
\text{$d_{x^2-y^2}$-wave:}\; &\Delta(\boldsymbol{K}_\pm + \vk) = 3\Delta_t[1 + \frac{a}{2}(\i k_x \pm k_y)],\notag\\
\text{$d_{xy}$-wave:}\; &\Delta(\boldsymbol{K}_\pm + \vk) = \sqrt{3}\Delta_t[\pm\i + \frac{a}{2}(-\i k_y \pm k_x)].
\end{align}
From Eq.~(\ref{eq:syms}), the classification of the different symmetries for nearest-neighbor pairing is far from obvious.
This is because in Eq.~(\ref{eq:syms}) they are expressed in reciprocal space where the kinetic energy is not diagonal. By 
diagonalizing the kinetic energy through an unitary transformation on $A$ and $B$, the correct symmetries will appear. 
These low-energy expansions are given in Sec.~\ref{sec:dwaves}.
Diagonalization of the Hamiltonian in Eq.~(\ref{eq:H}) yields the Dirac Bogoliubov-de Gennes (DBdG) equation which 
describes the quasiparticle excitations in the system. We find that the DBdG equation close to the Dirac points may 
be written as 
\begin{widetext}
\begin{align}\label{eq:eqham}
\begin{pmatrix}
-\mu & \varepsilon(\boldsymbol{K}_\pm + \vk) & \Delta_U & \Delta(\boldsymbol{K}_\pm + \vk) \\
\varepsilon^\dag(\boldsymbol{K}_\pm + \vk) & -\mu & \Delta(\boldsymbol{K}_\mp - \vk) & \Delta_U \\
\Delta_U^\dag & \Delta^\dag(\boldsymbol{K}_\mp -\vk) & \mu & -\varepsilon^\dag(\boldsymbol{K}_\mp - \vk) \\
\Delta^\dag(\boldsymbol{K}_\pm + \vk) & \Delta_U^\dag & -\varepsilon(\boldsymbol{K}_\mp -\vk) & \mu\\
\end{pmatrix}\psi = E\psi.
\end{align}
Due to the valley degeneracy, it suffices to consider only the Dirac point $K_+$. For concreteness, we include on-site and 
extended $s$-wave symmetry superconducting pairing below. It should be noted that the superconducting pairing potential couples 
electron and hole excitations between the two valleys $K_+$ and $K_-$ in Eq. (\ref{eq:eqham}), as required by time-reversal 
symmetry (within a single valley, time-reversal symmetry is broken). In this case, we find that
\begin{align}
\begin{pmatrix}\label{eq:bdg_swave}
-\mu & v_F(-\i k_x+k_y) & \Delta_U & 3\Delta_ta(\i k_x-k_y)/2 \\
v_F(\i k_x+k_y) & -\mu & -3\Delta_ta(\i k_x+k_y)/2 & \Delta_U \\
\Delta_U^\dag & 3\Delta_t^\dag a(\i k_x-k_y)/2 & \mu & -v_F(-\i k_x+k_y)\\
-3\Delta_t^\dag a(\i k_x+k_y)/2 & \Delta_U^\dag & -v_F(\i k_x+k_y) & \mu\\
\end{pmatrix}
\psi = E\psi.
\end{align}
\end{widetext}
Note that normal-state (non-superconducting) contribution to the above is slightly different from the usual Dirac equation $-\mu\sigma_0 + v_F\boldsymbol{p}\cdot\boldsymbol{\sigma}$. In fact, the upper-left $2\times2$ matrix of Eq. (\ref{eq:bdg_swave}) may be written as $-\mu\sigma_0 + v_F(k_x\sigma_y + k_y\sigma_x)$. The reason for this discrepancy is that the exact form of the Hamiltonian depends on the choice of nearest-neighbor vectors $\boldsymbol{a}_j,\; j\in\{1,2,3\}$, that were introduced previously. The physics must clearly remain completely unchanged regardless of the choice of $\boldsymbol{a}_j$. However, to facilitate comparison with previous work in the literature we revert to a choice that yields a Dirac-like Hamiltonian for the normal-state. This is simply accomplished by switching the coordinate system chosen originally, i.e. $k_x\leftrightarrow k_y$. Performing this substitution in Eq. (\ref{eq:bdg_swave}) yields the desired form of the normal-state Hamiltonian.
\par
The strategy for calculating the Josephson current in the junction is to first obtain the energy spectrum for the Andreev bound states in the normal region of graphene. This is done by matching the wavefunctions at the two SN interfaces, and then solving for the allowed energy states. Explicitly, the boundary conditions dictate that $\Psi_L|_{x=0} = \Psi_N|_{x=0}$ and $\Psi_R|_{x=L} = \Psi_N|_{x=L}$, where $L$ is the length of the N region, i.e.~the junction length and
\begin{align}
\Psi_N &= t_1\psi^\text{e}_+ + t_2\psi^\text{e}_- + t_3\psi^\text{h}_+ + t_4\psi^\text{h}_-,\notag\\
\Psi_L &=  t_L^\text{e}\Psi^\text{e}_- + t_L^\text{h}\Psi^\text{h}_-,\; \Psi_R =  t_R^\text{e}\Psi^\text{e}_+ + t_R^\text{h}\Psi^\text{h}_+.
\end{align}
We allow for the chemical potential to be different in the S and N regions.
Finally, note that the subscript $\pm$ on the wavefunctions in the normal region indicates the direction of their group velocity, which \textit{in general} is different from the direction of momentum. Consequently, although the Andreev-reflected hole wavefunction carries a subscript "$-$" above, one should keep in mind that for normal Andreev reflection, the direction of momentum is opposite to the group velocity for the hole.
\par
The Josephson current is computed via the usual energy-current relation summed over projections of all paths perpendicular to the tunneling barrier
\cite{beenakker_prl_91} 
\begin{align}\label{eq:jos}
I_J(\Delta\phi) &= \frac{4e}{\hbar} \sum_i \int^{\pi/2}_{-\pi/2} \frac{\text{d}\gamma \cos\gamma}{f^{-1}[\varepsilon_i(\Delta\phi)]} \frac{\text{d}\varepsilon_i(\Delta\phi)}{\text{d} \Delta\phi} ,
\end{align}
where $\varepsilon_i(\Delta\phi)$ are the Andreev bound states carrying the current in the N region, and 
$\Delta\phi=\phi_\text{R} -\phi_\text{L}$ is the macroscopic phase difference between the superconductors. The integration 
over angles $\gamma$ takes into account all possible trajectories and $f(x)$ is the Fermi-Dirac distribution function. We 
define the critical supercurrent as $I_c = |\text{max}\{I_J(\Delta\phi)\}|$, and note that the factor of 4 in front of the summation in Eq. (\ref{eq:jos}) is due to the spin-valley degeneracy. 
The formula for the Josephson current disregards the contribution from supergap states, which is allowed as long as $L/\xi\ll 1$, i.e.~a short junction. 
\subsection{Self-consistent numerical treatment}
A self-consistent numerical treatment allows for spatially varying order parameters, $\Delta_U$ and $\DeltaJ$, and will thus directly capture the proximity effect inside the junction through the depletion of pair amplitude in the superconductor near the interface and the induction of pair amplitude into the normal region. From this it is also possible to explicitly calculate the full Josephson current without any restrictions to small junctions as is the limitation for the analytical treatment with Andreev bound states. Naturally, the local density of states (LDOS) will also be readily available.
Here, we will use the tight-binding Bogoliubov-de Gennes (TB BdG) formalism, which allows for a self-consistent solution of the order parameters as a function of position. The procedure has been outlined in detail in Refs.~[\onlinecite{black-schaffer_prb_08}] and [\onlinecite{black-schaffer_prb_09}] and we will here only outline the essentials in order to connect to the analytical treatment and interpret the results.
\par
As in the case of the analytical treatment above of a graphene SNS Josephson junction we only want to model the actual graphene sheet. Therefore, we have to capture the effect of the superconducting contacts deposited on top of the sheet by some effective parameters. In the analytical treatment this was simply done by assuming that the contacts induce constant order parameters, or gaps, $\Delta_U$ and $\DeltaJ$, in the S regions.  For spatially varying order parameters we need to go beyond that approximation and we will instead model the effect of the the superconducting contacts by using effective pairing potentials which are only nonzero in the S regions of the graphene sheet. For $s$-wave contacts the pairing correlations induced into the graphene are modeled by a simple attractive Hubbard $U$-term. The nearest neighbor SB pairing can in the same way be produced by an effective SB potential $J$ where the effective coupling is given by a $J {\bf S_i\cdot S_j}$ term between nearest neighbors.\cite{black-schaffer_prb_09}
Thus, the starting point for a TB BdG treatment is the following effective Hamiltonian \cite{black-schaffer_prb_09}
\begin{align}
\label{eq:Heff}
H_{\rm eff} &= -t\sum_{ij\sigma} (A_{i\sigma}^\dag B_{i+\boldsymbol{a}_j,\sigma} + \text{h.c}) \notag\\
&- \sum_{i\sigma} \mu(i) (A_{\i\sigma}^\dag A_{\i\sigma} + B_{i,\sigma}^\dag B_{i,\sigma}) \notag\\
&- \sum_i U(i) (A_{i\uparrow}^\dag A_{i\uparrow} A_{i\downarrow}^\dag A_{i\downarrow}+ B_{i\uparrow}^\dag B_{i\uparrow} B_{i\downarrow}^\dag B_{i \downarrow})  \notag\\
&- \sum_{ij} 2J(i) F_{ij}^\dag F_{ij},
\end{align}
 where
\begin{align}
\label{eq:Fij}
F_{ij}^\dag = \frac{1}{\sqrt{2}}(A_{i\uparrow}^\dag B_{i+\boldsymbol{a}_j,\downarrow}^\dag - A_{i\downarrow}^\dag B_{i+\boldsymbol{a}_j,\uparrow}^\dag)
\end{align}
 is the nearest neighbor SB creation operator.
 With a simple Hartree-Fock-Bogoliubov mean-field approximation Eq.~(\ref{eq:Heff}) can be transformed into
 Eq.~(\ref{eq:Hana}), but now with the position dependent order parameters 
\begin{align}
\label{eq:Ds}
 \Delta_U(i) & = -U(i)\frac{\langle A_{i\downarrow} A_{i\uparrow}\rangle + \langle B_{i\downarrow} B_{i\uparrow}\rangle}{2} \\
 \Delta_{J\boldsymbol{a}_j}(i) & = -J(i)\langle A_{i\downarrow} B_{i+\boldsymbol{a}_j \uparrow} - A_{i\uparrow} B_{i+\boldsymbol{a}_j\downarrow}\rangle.
\end{align}
A standard TB BdG formulation of this mean-field Hamiltonian, Eq.~(\ref{eq:Hana}), will result in a $4N \times 4N$ eigenvalue problem, where $N$ is the number of unit cells in the whole junction, and position-dependent, BCS self-consistency equations for $\Delta_U(i)$ and $\DeltaJ(i)$. By starting with an initial guess for the order parameters $\Delta_U(i)$ and $\DeltaJ(i)$, then solving the eigenvalue problem for these values, and finally using the BCS self-consistency equations, we can compute new values for the order parameters and continue the process until self-consistency is reached. For a specific symmetry of the SB pairing contacts, we simply restrict $\DeltaJ$ to that particular symmetry.
Since the order parameters are by definition zero in the N region where $U$ and $J$ are zero, the relevant parameters to describe the proximity effect inside the junction are instead the pairing amplitudes $F_U(i) = -\Delta_U(i)/U(i)$ and $F_{J\boldsymbol{a}_j}(i) = -\Delta_{J\boldsymbol{a}_j}(i)/[\sqrt{2} J(i)]$.
The other significant quantity to study in a SNS junction is the Josephson current. We calculate this by fixing the phase of the order parameters in the very end of the contacts and then solving self-consistently in the rest of the sample. The Josephson current can be calculated relatively straight-forwardly using the continuity equation and the Heisenberg equation for the electron density, see Ref.~[\onlinecite{black-schaffer_prb_08}] for further details. It should be noted that in this approach the phase of the order parameter is allowed to vary even in  the S regions, except at the very end of the contacts. This ensures true bulk-like conditions and gives a consistent Josephson current throughout the structure. However, it has the side-effect that the phase difference $\Delta \phi$ over the junction itself will always be less than $\pi$, since that is the largest phase difference we can apply across the whole structure, but part of this drop will necessarily take place in S if the current is non-zero. While this appears as a numerical artifact in this context, it is in fact closely related to the physical $2\pi$ phase-slip process in Josephson junctions (see e.g.~Ref.~[\onlinecite{Tinkhambook}]).

\subsubsection{Simulation details}
Since the $\DeltaJ$ superconducting state corresponds to pairing along the nearest neighbor bonds on the bipartite honeycomb lattice there will be a directional dependence for this state. In particular, different interfaces will behave differently. The two most common interfaces for the honeycomb lattice are the zigzag and the armchair interfaces, see Fig.~\ref{fig:graphene}, but any chiral interface is experimentally possible. 
Now, since the particular symmetries for the $\DeltaJ$ state, Eq.~(\ref{eq:sym}), are either fully symmetric or have a four-fold symmetry, the number of interface vs.~symmetry combinations can be reduced. Obviously, for the extended \swave\ the interface orientation does not matter. 
For the \dwave\ solutions, the solution for \daa\ on a zigzag interface should, at least to a good approximation, be the same as the \dbb\ on an armchair interface. But, as will be shown, the $d_{x^2-y^2}$- and \dbb s on the same interface behave significantly different. It is therefore of interest to study both symmetry solutions but it suffices to look at one interface. The analytic results in this article are calculated for an interface along the $y$-direction, i.e.~the zigzag interface, though the detailed shape of the  
interface is obviously irrelevant in a continuum model. For the numerical treatment, it turns out, however, that the \dbb\ is strongly suppressed at external edges, thus demanding bigger contact regions, so instead of studying the $d_{x^2-y^2}$- and \dbb s on the zigzag interface, we instead study the \daa\ state on both the zigzag and armchair interface, with the latter equivalent to the \dbb\ solution on the zigzag interface. 
In order to reduce the computational cost we consider only clean, smooth interfaces and Fourier transform in the direction parallel to the interface, thus significantly reducing the number of unit cells $N$. From an experimental point of view this entails studying junctions with an infinite width, though the periodicity of the solution is limited to one unit cell.
\par
The physical input parameters in the TB BdG treatment are the on-site pairing potential $U$ for conventional \swave\ contacts, the SB pairing potential $J$ for unconventional contacts, the effective potential $\mu$ in S and N, the length $L$ of N, and temperature. 
For conventional \swave\ superconducting contacts we choose the following setup: $U({\rm S}) = 3.4$~eV $= 1.36t$, $\mu_\text{S} = 1.5$~eV = 0.6$t$. This leads to $\Delta_U = 0.1$~eV in the bulk which corresponds to a  
superconducting coherence length $\xi = \hbar v_F / \Delta_U \approx  
50$~\AA\ which is 25 unit cells in the zigzag direction and 40~unit cells in the armchair direction. These values satisfy $\lambda_F(S) \ll \xi$, allow us to numerically investigate both the $L<\xi$ and $L>\xi$ cases, and coincide with previous work.\cite{black-schaffer_prb_08} They are, however, quite large values for a realistic situation but a smaller superconducting gap leads to a slower convergence rate and also a need for a larger system making calculations less feasible. We have checked our key results for smaller $U$ and found no significant difference. 
\par
In order to be able to compare the results between the conventional and unconventional contacts we need to use the same strength superconductor. This is not completely trivial since the unconventional contacts have an energy gap that varies on the Fermi surface. For the proximity effect in a SNS junction the relevant scale is the coherence length $\xi$ since this parameter will determine the superconducting decay length. We therefore choose the effective pairing potentials such that $\xi$ is unchanged between the different symmetries. For the unconventional superconducting contacts we calculate $\xi = \hbar v_F/\Delta_F$, with $\Delta_F = \sqrt{\langle \Delta_\vk^2 \rangle_{\rm FS}}$ being the average of the energy gap $\Delta_\vk$ over the Fermi surface.\cite{benfatto_prb_02} 
Since the doping level is very high in the contacts, we can approximate the band structure to only consist of one band and the energy gap parameter for the unconventional contacts then simplifies to\cite{black-schaffer_prb_07}
\begin{align}
\label{eq:energygap}
\Delta_\vk = \sum_j \Delta_{J\boldsymbol{a}_j} \cos(\vk \cdot \boldsymbol{a}_j - \varphi_\vk), 
\end{align}
where $\varphi_\vk = {\rm arg}(\sum_j {\rm e}^{i \vk \cdot \boldsymbol{a}_j})$.
Using this we find $\Delta_F$(\dwave) $= \frac{\sqrt{3}}{2}\Delta_J$ where $\Delta_J$ 
is the norm of $\DeltaJ$.  
Then finally, by comparing the coherence lengths, we can set $J({\rm S}) = 2.45$~eV for a straightforward comparison with the conventional \swave\ contacts. 
Please note that in the analytical formalism we need to choose $\Delta_t$ different for all three symmetries in order to have the same $\xi$. For example, if $\Delta_U = \Delta_0$ then $\Delta_t$(ext.~$s$) $= \frac{2}{3}\Delta_0$, $\Delta_t(d_{x^2-y^2})= \frac{\sqrt{2}}{3}\Delta_0$, and $\Delta_t(d_{xy})= \frac{\sqrt{2}}{\sqrt{3}}\Delta_0$, respectively.
\par
We have for simplicity assumed that the doping profile changes abruptly from $\mu_\text{S}$ to $\mu_\text{N}$ at the interface. This is the same approximation as in the analytical solution and will therefore provide the most accurate comparison. We have studied junctions ranging from the undoped regime, $\mu_N= 0$~eV, to moderately doped, $\mu_\text{N} = 0.7$~eV, to the case of no Fermi wavevector mismatch (FVM), i.e.~$\mu_\text{N} = 1.5$~eV.
\par
In terms of $L$, we have studied zigzag junctions with $L$ = 2-60 unit cells (1 unit cell = 2.13 \AA) and  armchair junctions with the corresponding number of cells (1 unit cell = 1.23 \AA).
The temperature was chosen to be $T = 10$~K throughout the work, which in comparison to $T_c$ in the S regions is effectively zero temperature.
The accuracy of the solution is determined by the choice of termination criterion for the self-consistency step, the number of $k$-points used in the Fourier transform, the size of S and the region in S where the phase of the order paramter is kept constant, all of which have been tested thoroughly.
\section{Results and Discussion}\label{sec:results}

\subsection{Analytical results}\label{sec:results_analytical}
In order to investigate how the unconventional pairing near the Dirac points affects the Josephson current, we begin by obtaining an analytical solution for the most possible general case. In a realistic situation, there could be both on-site and nearest-neighbor interactions, thus giving rise to both isotropic $s$-wave pairing and one of the order parameters given in Eq. (\ref{eq:syms}). 

\subsubsection{Extended $s$-wave}
We first consider the extended $s$-wave case, which gives rise to an effective $p+\i p$-wave order parameter near the Dirac points. The BdG-equation then reads
\begin{align}
\begin{pmatrix}\label{eq:bdg}
-\mu & p\e{-\i\theta} & \Delta_U & \Delta_T\e{-\i\theta} \\
p\e{\i\theta} & -\mu & \Delta_T\e{\i\theta}& \Delta_U \\
\Delta_U^\dag & \Delta_T^\dag \e{-\i\theta} & \mu & -p\e{-\i\theta}\\
\Delta_T^\dag\e{\i\theta} & \Delta_U^\dag & -p\e{\i\theta} & \mu\\
\end{pmatrix}
\psi = \varepsilon\psi.
\end{align} 
We here use units $\hbar=v_F=1$ and have defined $p_x\pm\i p_y \equiv p\e{\pm\i\theta}$ and $\Delta_T = -3p\Delta_t a/2$. The above matrix is Hermitian as required and may be diagonalized to yield the eigenvalues
\begin{align}\label{eq:dispersion}
\varepsilon_{\alpha\beta} = \alpha\sqrt{(\beta p -\mu)^2 + |\Delta_U + \beta \Delta_T|^2},\; \alpha,\beta=\pm.
\end{align}
Here, $\beta=\pm1$ denotes the conduction and valence band while $\alpha=\pm1$ distinguishes between electron-like and hole-like excitations. The eigenfunctions may then be constructed in the superconducting regions. We here only consider positive excitation energies $\varepsilon\geq 0$, and impose the mean-field restriction that the chemical potential $\mu_\text{S}$ in the superconducting regions must be much larger than the superconducting gap, i.e. $\mu_\text{S}\gg\Delta$. In this case, only the conduction band $\beta=1$ partakes in the low-energy scattering processes. We define $[v(p)]^2 = 1 - [u(p)]^2$ with
\begin{align}
u(p) = \sqrt{\frac{1}{2}\Big(1 + \frac{\sqrt{\varepsilon^2 - |\Delta_U + \Delta_T(p)|^2}}{\varepsilon}\Big)}.
\end{align}
From Eq.~(\ref{eq:dispersion}), one can then solve for the wavevector as follows ($\alpha=1$ since $\varepsilon \geq 0$):
\begin{align}
q_\text{e,h}= \frac{\mu_\text{S} -\Delta_U\tilde{\Delta}_T\pm \sqrt{\varepsilon^2(1+\tilde{\Delta}_T) - (\mu\tilde{\Delta}_T+\Delta_U)^2}}{1+\tilde{\Delta}_T^2},
\end{align}
where $\tilde{\Delta}_T = \Delta_T/p = -3\Delta_ta/2$ and the scattering angles are obtained by means of translational invariance in the $y$-direction:
\begin{align}
q_i\sin\theta_i = p_e\sin\theta,\; i=\text{e,h}
\end{align}
We also define $\Delta\mu = \mu_S-\mu_N$ and
\begin{align}
p_e = \varepsilon+\mu_N,\;\; p_h = \varepsilon-\mu_N,\;\; p_h\sin\theta_A = p_e\sin\theta,
\end{align}
related to the quasiparticle momenta in the normal graphene region. Here, $\Delta\mu$ is the difference in the local chemical potential between the S and N region, which may be experimentally controlled by means of a gate voltage on top of the normal graphene segment.
\par
Finally, we are able to write down the wavefunctions in the three regions shown in Fig.~\ref{fig:model}. We remind the reader that the effective pairing near the Dirac points is a mixture of isotropic $s$-wave and $p+\i p$-wave. However, note that while the extended $s$-wave in this picture corresponds to the even and complex $p+\i p$-wave symmetry, it breaks neither time-reversal symmetry nor is it a spin-triplet. The deceptive appearance is due to the fact that the honeycomb lattice has two distinct Fermi surfaces at all doping levels considered here and when considered together, the spin-singlet character and TRS invariance is preserved as it should be for an extended $s$-wave. Note that in what follows, we will assume that the $s$- and $p+\i p$-wave superconducting order parameters are phase-locked, i.e.~they are characterized by the same broken U(1) gauge phase. For $x<0$, we have
\begin{align}\label{eq:w1}
\Psi_{S,L} &= t_{L}^e\begin{pmatrix}
u(q_e)\\
u(q_e)\e{\i(\pi-\theta_e)}\\
v(q_e)\e{-\i\phi_L}\\
v(q_e)\e{\i(\pi-\theta_e-\phi_L)}\\
\end{pmatrix}
\e{-\i q_e\cos\theta_e x} \notag\\
&+ 
t_{L}^h\begin{pmatrix}
v(q_h)\\
v(q_h)\e{\i\theta_h}\\
u(q_h)\e{-\i\phi_L}\\
u(q_h)\e{\i(\theta_h -\phi_L)}\\
\end{pmatrix}
\e{\i q_h\cos\theta_h x},
\end{align}
\begin{align}\label{eq:w2}
\Psi_N &= a\begin{pmatrix}
1\\
\e{\i\theta}\\
0\\
0\\
\end{pmatrix}
\e{\i p_e\cos\theta x} + b\begin{pmatrix}
1\\
-\e{-\i\theta}\\
0\\
0\\
\end{pmatrix} 
\e{-\i p_e\cos\theta x} \notag\\
&+ c \begin{pmatrix}
0\\
0\\
1\\
\e{-\i\theta_A}
\end{pmatrix} 
\e{-\i p_h\cos\theta_A x} + d \begin{pmatrix}
0\\
0\\
1\\
-\e{\i\theta_A}
\end{pmatrix}
\e{\i p_h\cos\theta x},
\end{align}
\begin{align}\label{eq:w3}
\Psi_{S,R} &= t_{R}^e\begin{pmatrix}
u(q_e)\\
u(q_e)\e{\i\theta_e}\\
v(q_e)\e{-\i\phi_R}\\
v(q_e)\e{\i(\theta_e-\phi_R)}\\
\end{pmatrix}
\e{\i q_e\cos\theta_e x} \notag\\
&+ 
t_{R}^h\begin{pmatrix}
v(q_h)\\
v(q_h)\e{\i(\pi-\theta_h)}\\
u(q_h)\e{-\i\phi_R}\\
u(q_h)\e{\i(\pi-\theta_h-\phi_R)}\\
\end{pmatrix}
\e{-\i q_h\cos\theta_h x},
\end{align}
where $\phi_{L,R}$ is the superconducting phase on the left/right side of the normal region, associated with the broken U(1) symmetry in the superconducting state. The macroscopic phase difference is defined as $\Delta\phi = \phi_R-\phi_L$.
\par
In order to proceed with an analytical treatment, we observe that $q_{e,h}=\mu_\text{S}$ under the assumption that $\mu_\text{S}\gg \{\Delta_U,\Delta_T\}$. Moreover, the direction of the momentum that enters the argument of the coherence functions $\{u,v\}$ in the wavefunctions of Eqs.~(\ref{eq:w1}-\ref{eq:w3}) is of no significance since only the absolute value of the momentum enters $\Delta_T$. The directional dependence has been separated out into the $\e{\i\theta}$ factors of the off-diagonal elements in Eq.~(\ref{eq:bdg}). Thus, the problem effectively becomes equivalent to that of a conventional $s$-wave superconductor with gap $\Delta_0=|\Delta_U -3\mu_\text{S}\Delta_t a/2|$. So, while we have included both onsite and nearest neighbor interactions, thus giving rise to a combination of $s$-wave and $p+\i p$-wave pairing, the results would have been identical had we chosen only on-site or nearest neighbor interaction, as long as the nearest neighbor interaction gives rise to the extended $s$-wave symmetry in Eq.~(\ref{eq:syms}). 
This conclusion is supported by the findings of Jiang \etal,\cite{jiang_prb_08} who found that the Andreev conductance of a SN junction was identical for $s$-wave pairing and the extended $s$-wave bond pairing (see their Fig.~3c).

We now obtain the following energies for the Andreev bound states in the normal region:
\begin{align}
\varepsilon_\pm &= \pm |\Delta_U -3\mu_\text{S}\Delta_t a/2|\sqrt{1 - \zeta(\gamma)\sin^2(\Delta\phi/2)},\notag\\
\zeta(\gamma) &= \frac{\cos^2\gamma}{1-\sin^2\gamma\cos^2(P\cos\theta)},\; P = \mu_\text{N}L/v_F.
\end{align}
Note that $\theta=0$ if $\Delta\mu\gg\mu_\text{S}$ and $\theta=\gamma$ if $\Delta\mu=0$, where $\gamma=\theta_e$ is the angle of incidence of quasiparticles. In the former case, we regain the results of Maiti and Sengupta \cite{maiti_prb_07}, who studied the case of a thin and very strong barrier separating the two superconducting regions. In the case $\Delta\mu=0$, we have no FVM between the graphene regions ($\mu_\text{S}=\mu_\text{N}=\mu$), and the results change accordingly. In this case the normalized Josephson current at zero temperature becomes
\begin{align}
I_J/I_0 = \int^{\pi/2}_{-\pi/2} \frac{\text{d}\gamma \zeta(\gamma) \cos\gamma\sin\Delta\phi}{\sqrt{1 - \zeta(\gamma)\sin^2(\Delta\phi/2)}},
\end{align}
where $I_0=e\Delta_0/\hbar$. In the following, we will study the dependence of the Josephson current on the (\textit{i}) phase difference, (\textit{ii}) doping level and length of the junction, and (\textit{iii}) the temperature-dependence. Since we consider the regime $\mu\gg\{\Delta_U,\Delta_T\}$, which means that $\mu$ typically could lie in the range $10-100$ meV, it is reasonable to expect that the role played by charge inhomogeneities such as electron-hole puddles may be disregarded in the main approximation. To be more specific, the local variations of the Fermi level $\delta\mu$ should be of little importance as long as they satisfy $\delta\mu \ll \mu$. Experimentally, one has estimated \cite{puddles} $\delta\mu\simeq 5$ meV, which places a restriction on the appropriate values of $\mu$.

\begin{figure}[b!]
\centering
\resizebox{0.48\textwidth}{!}{
\includegraphics{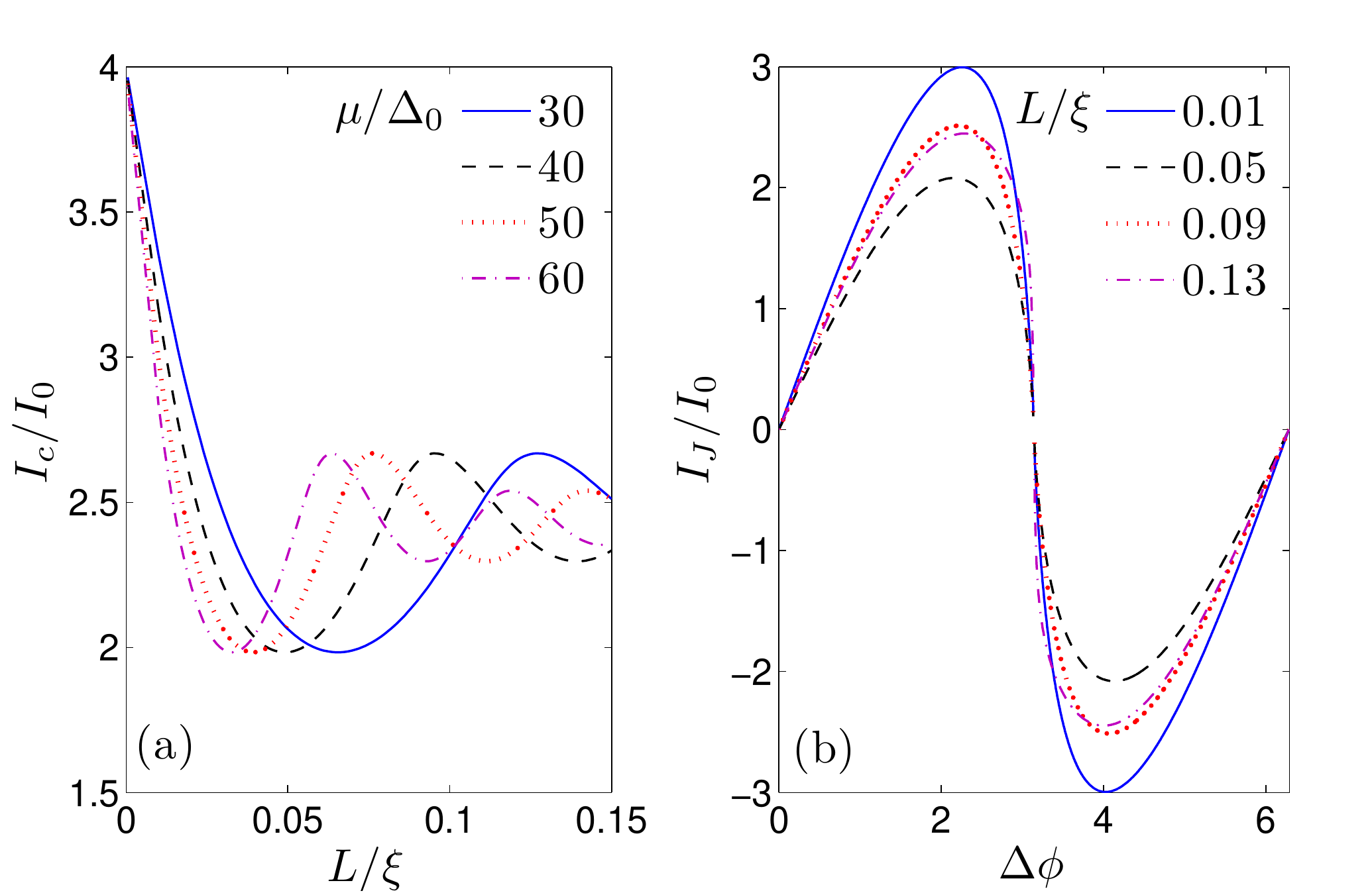}}
\caption{(Color online) Josephson current for the $s$-wave symmetry with no FVM between the S and N regions ($\mu_\text{S}=\mu_\text{N}=\mu$). (a) Plot of the length-dependence of the critical current. (b) Plot of the current-phase relationship for $\mu/\Delta_0=50$.}
\label{fig:swave_undoped}
\end{figure}
\begin{figure}[b!]
\centering
\resizebox{0.48\textwidth}{!}{
\includegraphics{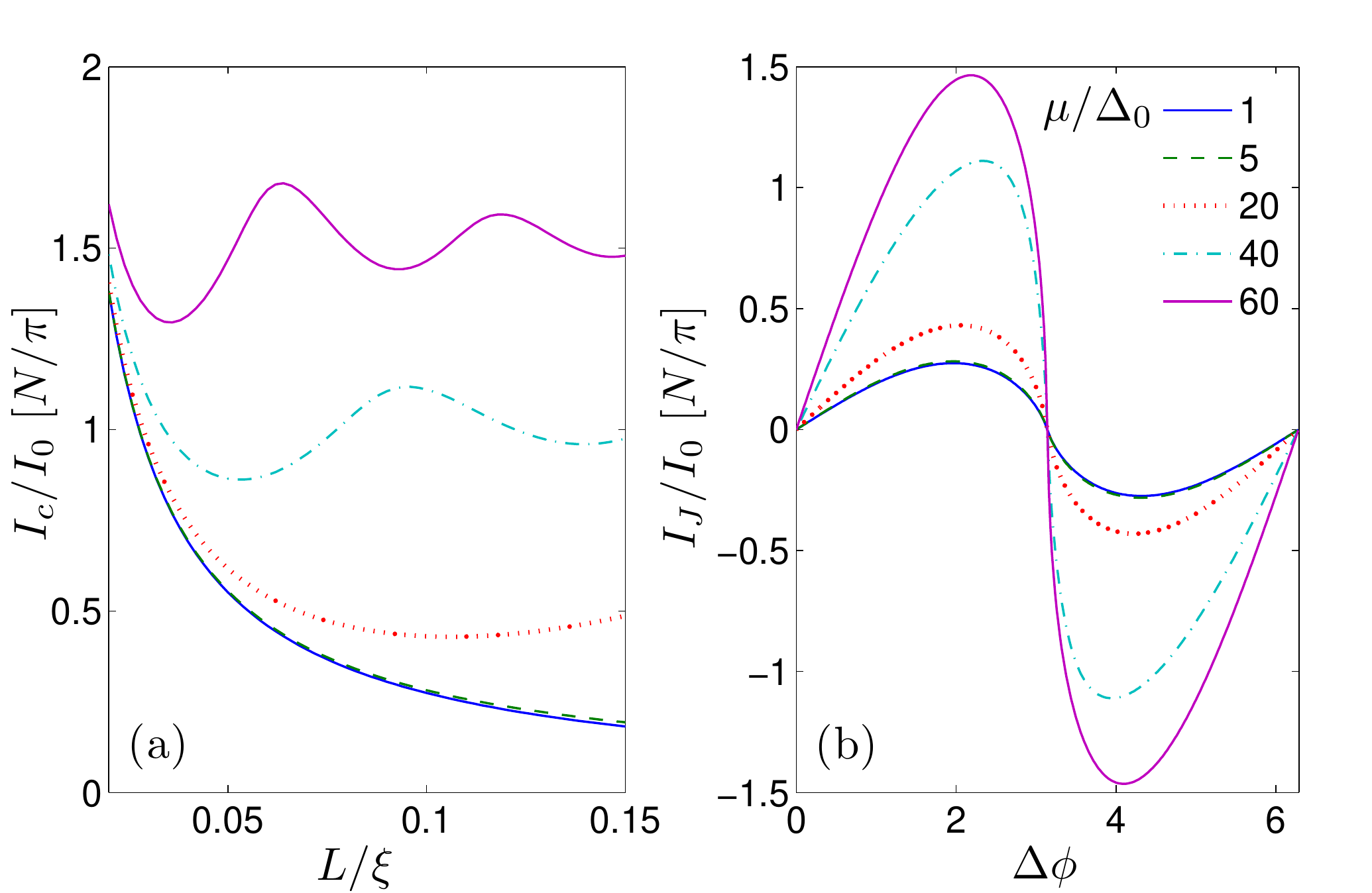}}
\caption{(Color online) Josephson current for the $s$-wave symmetry with a strong FVM between the S and N regions ($\mu_N=\mu$). (a) Plot of the length-dependence of the critical current. (b) Plot of the current-phase relationship for $L/\xi=0.1$.}
\label{fig:swave_doped}
\end{figure}

We now proceed to investigate the length-dependence of the critical current in Fig. \ref{fig:swave_undoped}(a). In all the plots presented in this section, the zero-temperature limit is assumed unless explicitly stated otherwise. To operate within a valid regime of parameters, we restrict our attention to $L/\xi \leq 0.15$, where $\xi=\hbar v_F/\Delta_0$ is the superconducting coherence length. As seen, the critical current displays an oscillating decay. This is a qualitatively new feature as compared to the results of Refs.~ [\onlinecite{titov_prb_06}] and [\onlinecite{maiti_prb_07}]. For a strongly doped graphene junction, but still with a FVM at the interfaces, Black-Schaffer \etal \cite{black-schaffer_prb_08} obtained numerically the length-dependence of the critical current. From Fig.~5(b) in that work, one may see a hint of oscillations in the numerical data and it is clear that the current does not decay monotonously. The same data, but with more data points, are reproduced in Fig~\ref{fig:IvsL}(b) and from this and Fig.~\ref{fig:swave_doped} it is clear that both the numerical and analytical data show an oscillatory dependence on $L$ in the strong doping regime including the case of no FVM.
\par
In Fig.~\ref{fig:swave_undoped}(b), we give the current-phase relationship for several values of the length of the normal-region, using $\mu/\Delta_0=50$, in the case when there is no FVM. As seen, the current-phase relationship deviates slightly from the usual sinusoidal form with the phase difference providing the critical current occurring at $\Delta\phi_c\in[\pi/2,\pi]$. Qualitatively, Fig.~\ref{fig:swave_undoped}(b) is in agreement with the numerical results shown in Fig.~4 of Ref.~[\onlinecite{black-schaffer_prb_08}]. Note that by lowering $\mu$ and $L$, the current-phase relationship tends towards the functional form $\sin(\Delta\phi/2)\text{sgn}\{\cos(\Delta\phi/2)\}$.

The analytical treatment above is valid for either $\Delta\mu=0$ (corresponding to zero FVM) or $\Delta\mu\gg\mu_\text{S}$ (corresponding to a barrier induced in the normal graphene region). Next, we consider a situation where the superconducting regions are strongly doped, while the normal region is only weakly or moderately doped. Since we have shown above that the effective $s$+$(p+\i p)$ wave pairing near the Dirac points is formally equivalent to the isotropic $s$-wave case, we can here use\cite{titov_prb_06}:
\begin{align}
I_J&= \frac{e\Delta_0}{\hbar} \sum_{n=0}^N \frac{\zeta_n\sin\Delta\phi}{\sqrt{1-\zeta_n\sin^2(\Delta\phi/2)}},\notag\\
\zeta_n &= k_n^2[k_n^2\cos^2(k_nL) + \mu_\text{N}^2\sin^2(k_n L)]^{-1},\notag\\
k_n &= \sqrt{\mu_\text{N}^2-q_n^2},\; q_n = (n+1/2)\pi/W.
\end{align}
Above, $W$ denotes the width of the graphene junction, which we assume satisfies $W\gg L$. We assume that the superconducting regions are heavily doped, such that $\mu_S\gg\mu_N$. In this case, the number of propagating modes in the superconducting region is $N=\mu_SW/\pi$. We now proceed to investigate how the critical current depends on the length $L$ of the junction. We choose $W/\xi=30$ and $\mu_S/\Delta_0=150$. The result is shown in Fig.~\ref{fig:swave_doped}(a). In general, the actual magnitude of the current decreases with decreasing $\mu_\text{N}$ since there are fewer propagating modes available when $\mu_N\to 0$. At weak doping $\mu_N/\Delta_0 \sim 1-10$, the current decays as $1.33I_0W/(\pi L)$, in agreement with the findings of Ref.~[\onlinecite{titov_prb_06}]. However, as the doping level increases, oscillations appear as a function of $L$. We also give the current-phase relation for the case with a strong FVM in Fig.~\ref{fig:swave_doped}(b), using $L/\xi=0.1$.
\subsubsection{$d_{x^2-y^2}$- and $d_{xy}$-wave}
\label{sec:dwaves}
As seen from Eqs. (\ref{eq:syms}), the situation becomes quite complicated when considering the $d$-wave symmetries that may arise upon considering nearest-neighbor pairing. At least, this is so when considering the DBdG equation in an atom-basis picture, as has been done in the literature up to now. However, by transforming the DBdG-equation to the band-structure basis (conduction and valence $\pi$-bands), i.e.~where the kinetic energy is diagonal, we are able to treat the $d$-wave symmetries analytically. This stems from the fact that the $d$-wave symmetries in this basis have a simply four-fold symmetry and their low energy expansions near the Dirac points result in even simpler $p$-waves.
Below, we sketch this transformation.
\par
\begin{figure}[htb]
\centering
\resizebox{0.45\textwidth}{!}{
\includegraphics{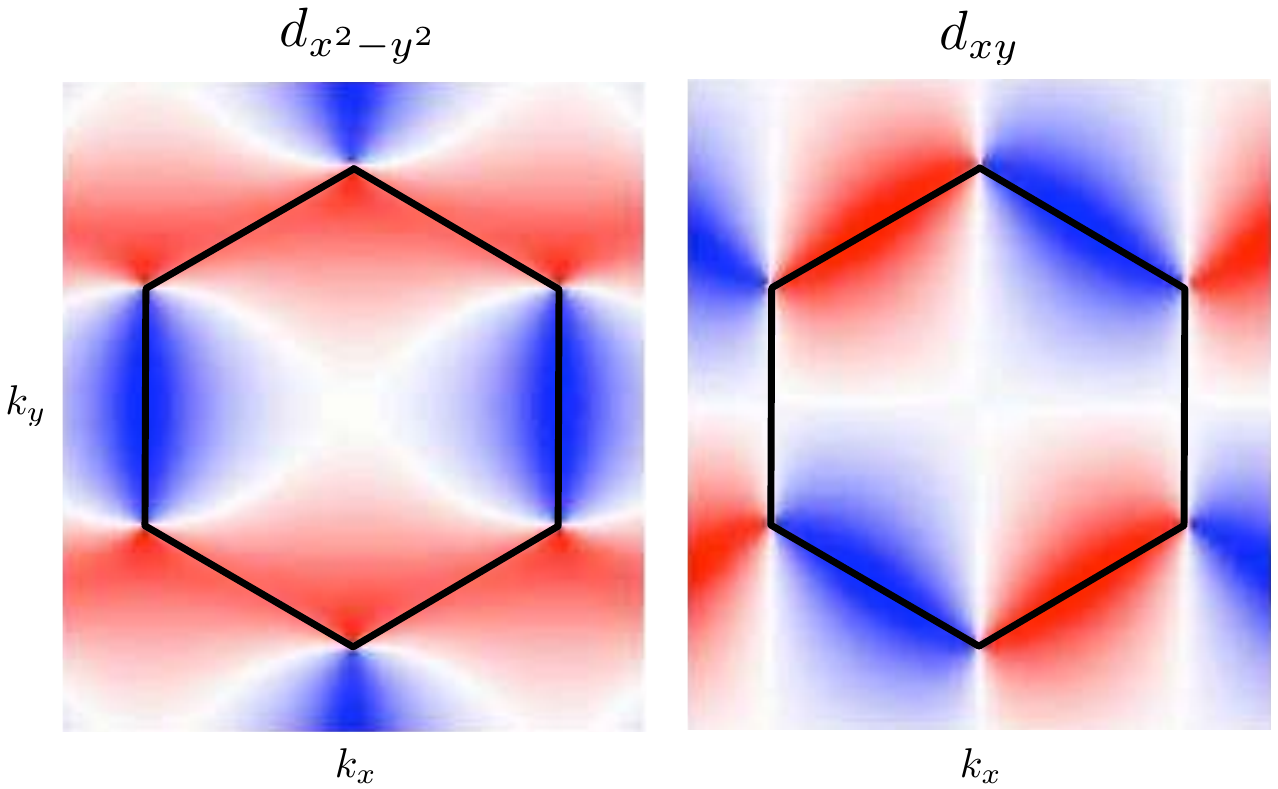}}
\caption{(Color online) Plot of the $\vq$-dependence of the two $d$-wave symmetry order parameters ($C_\vq$). Red color represent positive sign, blue color negative sign, with zero being white. The first Brillouin zone is marked with a black line. Near the Dirac points, $(q_x,q_y) = (0, \pm\frac{4\pi}{3\sqrt{3}a})$, at the zone corners, effective $p_y$- or $p_x$-wave symmetries emerge in the low-energy regime.}
\label{fig:gaps}
\end{figure}

Our starting point is Eq.~(7) in Ref.~[\onlinecite{black-schaffer_prb_07}], where the superconducting pairing is written in the band-picture as follows:
\begin{align}
H &= \sum_{\vq\sigma} [(t\epsilon_\vq - \mu)c_{\vq\sigma}^\dag c_{\vq\sigma} + (-t\epsilon_\vq - \mu) d_{\vq\sigma}^\dag d_{\vq\sigma}]\notag\\
&-\sum_{\vq,j} \boldsymbol{\Delta}_j [\cos(\vq\cdot\va_j - \varphi_\vq) (d_{\vq\uparrow} d_{-\vq\downarrow}^\dag -  c_{\vq\uparrow}^\dag c_{-\vq\downarrow}^\dag) \notag\\
&+\i\sin(\vq\cdot\va_j -\varphi_\vq)( c_{\vq\uparrow}^\dag d_{-\vq\downarrow}^\dag - d_{\vq\uparrow}^\dag c_{-\vq\downarrow}^\dag) + \text{H.c.}]
\end{align}
Here, we have defined 
\begin{align}
\label{eq:bandstr}
\epsilon_\vq = \Big|\sum_\va \e{\i\vq\va}\Big| = |\varepsilon_\vq|.
\end{align}
Introducing a basis in the band-picture $\psi_\vq^\dag = [c_{\vq\uparrow}^\dag, d_{\vq\uparrow}^\dag, c_{-\vq\downarrow}, d_{-\vq\downarrow}]$, we may write the Hamiltonian as $H = \sum_\vq \psi^\dag_\vq N_\vq \psi_\vq$  with
\begin{align}\label{eq:Nk}
N_\vq &= \begin{pmatrix}
t\epsilon_\vq - \mu & 0 & -C_\vq & \i S_\vq\\
0 & -t\epsilon_\vq - \mu & -\i S_\vq & C_\vq \\
-C_\vq^* & \i S_\vq^* & -t\epsilon_\vq + \mu & 0 \\
-\i S_\vq^* & C_\vq^* & 0 & t\epsilon_\vq +\mu\\
\end{pmatrix},
\end{align}
where we have defined
\begin{align}
\label{eq:Ck}
C_\vq &= \sum_\va \boldsymbol{\Delta}_j \cos(\vq\cdot\va_j - \varphi_\vq),\notag\\
 S_\vq &= \sum_\va \boldsymbol{\Delta}_j  \sin(\vq\cdot\va_j - \varphi_\vq),
\end{align}
Note that $N_\vq$ is Hermitian, so we know that its eigenvalues will be real. As seen, the quantities $\{C_\vq,S_\vq\}$ play the roles of the superconducting gaps for intra- and inter-band pairing, respectively. We are interested in the behaviour of $N_\vq$ near the Dirac points, effectively at the wavevector $\vq = \boldsymbol{K}_\pm + \vk$, where again $|\vk|\ll |\boldsymbol{K}_\pm|$, and $\boldsymbol{K}_\pm = \pm[0,4\pi/(3\sqrt{3}a)]$. By inserting this into $N_\vq$ and linearizing in $\vk$, we obtain for the $d_{xy}$-wave symmetry featuring $\boldsymbol{\Delta} = \Delta_t(0,1,-1)$ that
\begin{align}
C(\boldsymbol{K}_\pm + \vk) = \pm \frac{\Delta_t k_x}{\sqrt{3}|\vk|},\; S(\boldsymbol{K}_\pm + \vk) = -\frac{\Delta_t\sqrt{3}k_y}{|\boldsymbol{\vk}|}.
\end{align}
Effectively, this is a $p_x$-wave pairing for $C_\vk$ and $p_y$-wave pairing for $S_\vk$. In an equivalent manner, we obtain for the $d_{x^2-y^2}$-wave case of $\boldsymbol{\Delta} = \Delta_t(2,-1,-1)$ that 
\begin{align}
C(\boldsymbol{K}_\pm + \vk) = \mp \frac{3\Delta_t k_y}{|\vk|},\; S(\boldsymbol{K}_\pm + \vk) = -\frac{\Delta_tk_x}{2|\boldsymbol{\vk}|}.
\end{align}
Let us also briefly mention in passing that for the extended $s$-wave case, $\boldsymbol{\Delta} = \Delta_t(1,1,1)$, we obtain
\begin{align}
C(\boldsymbol{K}_\pm + \vk) = \frac{3\Delta_t |\boldsymbol{k}|a}{2},\; S(\boldsymbol{K}_\pm + \vk) = 0,
\end{align}
which is consistent with the result of the previous section, i.e.~the extended $s$-wave case gives rise to an effective conventional, fully gapped $s$-wave pairing near the Dirac points with a doping-dependent magnitude of the gap. The $\vq$-dependence in the entire BZ is shown for the $d$-wave gaps in Fig.~\ref{fig:gaps}.
\par
The question is now: have we managed to simplify the expressions compared to the atom-basis picture such that an analytical approach has been rendered viable in the $d$-wave case? At first sight, it appears that the situation is still rather complicated as there are two "gaps" in Eq.~(\ref{eq:Nk}), namely $C_\vk$ and $S_\vk$. However, upon diagonalizing $N_\vk$ to obtain its eigenvalues, we find 
\begin{align}\label{eq:eigenvalue}
E_\vk = \alpha \sqrt{ [t\epsilon_\vk +\beta \sqrt{(\mu^2+S_\vk^2)}]^2 + C_\vk^2},
\end{align}
where again $\alpha=\pm1$ refers to e-like and h-like particles, while $\beta=\pm1$ refers to the conduction and valence band. From Eq.~(\ref{eq:eigenvalue}), it is clear that $S_\vk$ simply renormalizes the chemical potential $\mu$ while $C_\vk$ is the true superconducting gap. Assuming here a doped situation where $\mu\gg\Delta_t$, we may certainly neglect $S_\vk$ compared to $\mu$, and we therefore set $S_\vk=0$ in Eq.~(\ref{eq:Nk}). The situation has now been considerably simplified. We are left with a single-gap superconductor with normal dispersion $t\epsilon_\vk$ and gap $C_\vk$ where the gap has a simple $p_x$- or $p_y$-wave symmetry, which allows us to continue analytically. 
\par
For the $p_x$-wave symmetry, the situation becomes qualitatively different from the conventional $s$-wave case since there are now zero-energy states (ZES) located near the interfaces. Since we are considering transport along the $x$-direction, the criterion for the existence of ZES is that the order parameter satisfies $\Delta(\theta) = -\Delta(\pi-\theta)$. Clearly, this is the case for a $p_x$-wave symmetry with $\Delta(\theta)\sim\cos\theta$, whereas the $p_y$-wave case does not host any ZES. Let us here consider the case of no FVM. In the absence of ZES, the Andreev-bound states in the normal region may be written quite generally as
\begin{align}\label{eq:boundpy}
\varepsilon_\pm &= \pm |\Delta(\gamma)|\sqrt{1 - \zeta(\gamma)\sin^2(\Delta\phi/2)},
\end{align}
where again $\zeta(\gamma)$ is the angularly resolved transmission coefficient in the normal-state and $\Delta\phi$ is the phase-difference between the superconducting regions. It is seen that Eq.~(\ref{eq:boundpy}) is formally equivalent to the Andreev bound-state spectrum for a conventional $s$-wave symmetry, with the only difference that the gap now has an angular dependence, i.e. $\Delta = \Delta(\gamma)$. As a result, one would expect qualitatively the same results for the Josephson current when comparing the $p_y$-wave case with the $s$-wave case. Quantitatively, the magnitude of the current would be reduced due to angular averaging over the gap.
\par
\begin{figure}[h!]
\centering
\resizebox{0.48\textwidth}{!}{
\includegraphics{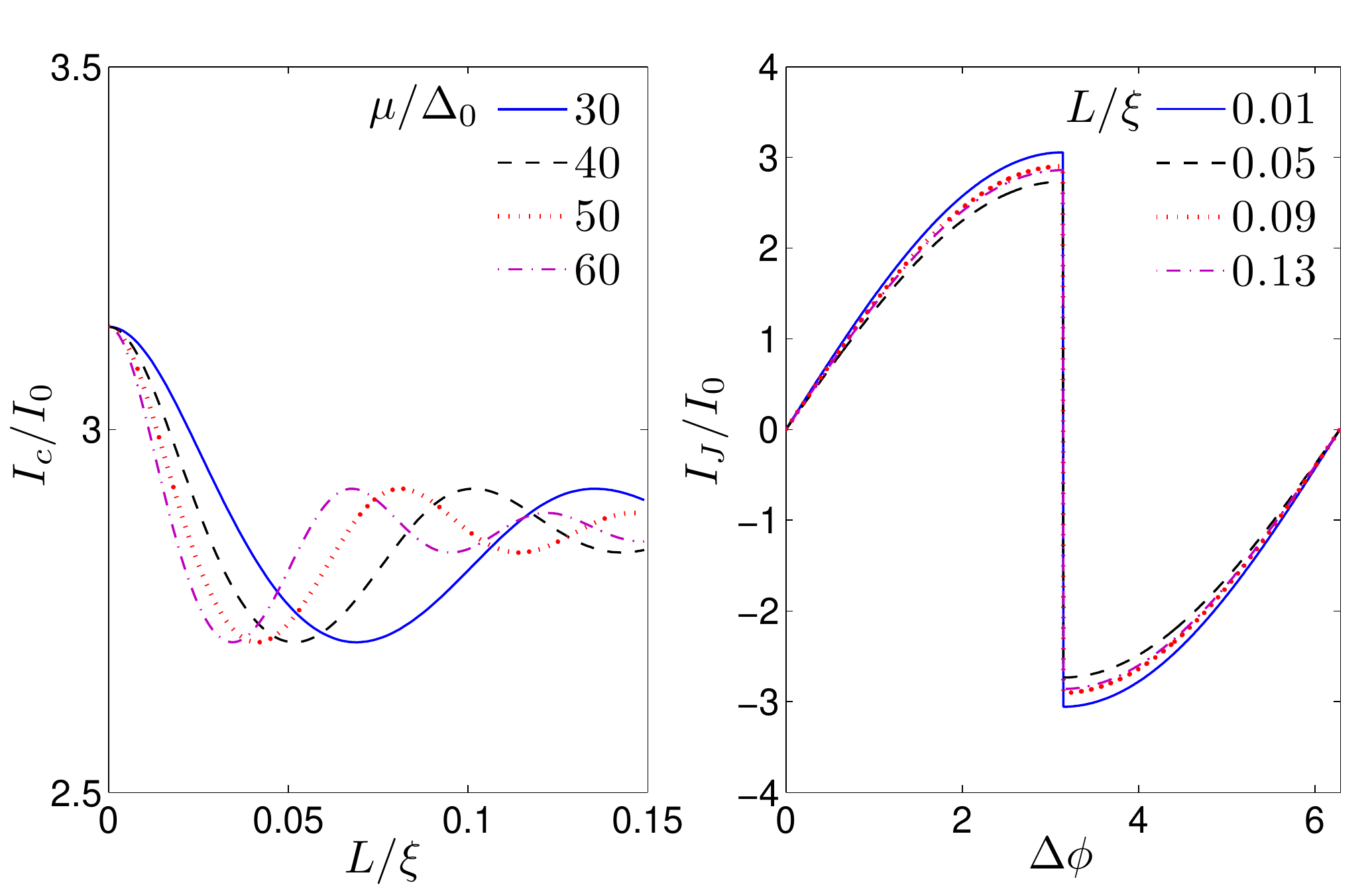}}
\caption{(Color online) Josephson current for the $d_{xy}$-wave symmetry (giving rise to an effective $p_x$-wave symmetry at the Dirac points) with no FVM between the S and N regions. (a) Plot of the length-dependence of the critical current. (b) Plot of the current-phase relationship for $\mu/\Delta_0=50$.}
\label{fig:dwave_undoped}
\end{figure}

We now consider the $p_x$-wave symmetry featuring ZES, and one then in general finds\cite{lofwander_ss_01}
\begin{align}\label{eq:boundstate_ZES}
\varepsilon_\pm &= \pm |\Delta(\gamma)| \sqrt{\zeta(\gamma)} \cos(\Delta\phi/2), 
\end{align}
where, $\Delta(\gamma) = \Delta_0\cos\gamma$. The normalized Josephson current then becomes
\begin{align}
I_J/I_0 &= 2\mathcal{J}\sin(\Delta\phi/2)\text{sgn}\{\cos(\Delta\phi/2)\},\notag\\
\mathcal{J} &=  \int^{\pi/2}_{-\pi/2} \text{d}\gamma\cos^2\gamma\sqrt{\zeta(\gamma)},\notag\\
\zeta(\gamma) &= \frac{\cos^2\gamma}{1-\sin^2\gamma\cos^2(P\cos\gamma)},\; P=\mu L/v_F.
\end{align}
We proceed to discuss the length-dependence and phase-dependence of this Josepshon current, and especially investigate how it differs from the conventional $s$-wave case where there are no ZES. As seen in Fig.~\ref{fig:dwave_undoped}(a), the current still displays oscillations as a function of the width $L$ in the strongly doped case $\mu\gg\Delta_0$, but the oscillation-amplitude is considerably smaller than for the $s$-wave symmetries. In Fig. \ref{fig:dwave_undoped}(b), we give the current-phase relationship for the Josephson junction. As seen, there is now an abrupt crossover at $\Delta\phi=\pi$ which should be contrasted with the smooth behavior in the $s$-wave case shown in Fig.~\ref{fig:swave_undoped}(b). An interesting aspect is that in the present case of $d$-wave symmetry, the junction energy is minimized at $\Delta\phi=\pi$, while in the $s$-wave case the minimum of free energy occurs at a phase difference $\phi_0$ which lies between 0 and $\pi$. 

\begin{figure}[h!]
\centering
\resizebox{0.48\textwidth}{!}{
\includegraphics{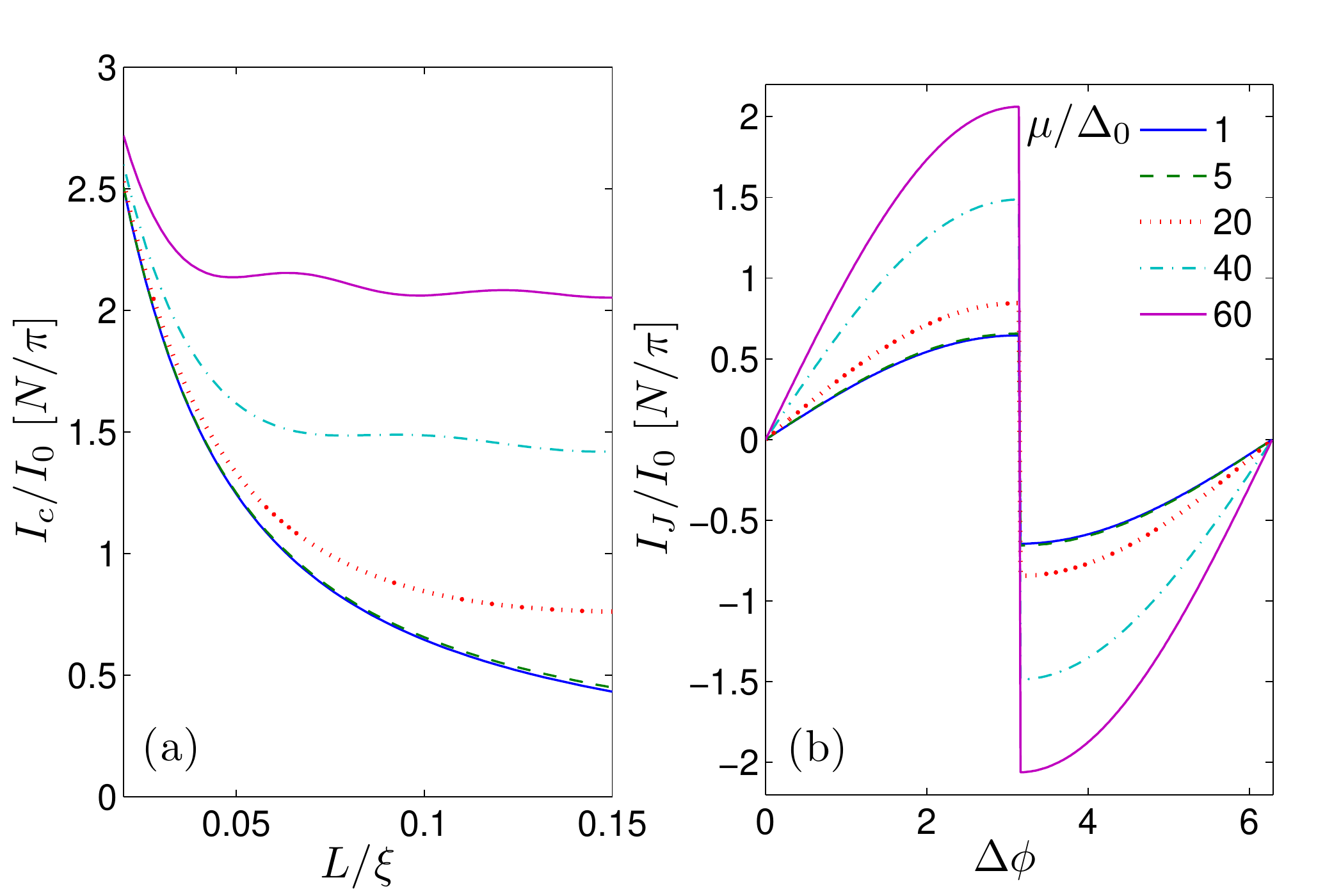}}
\caption{(Color online) Josephson current for the $d_{xy}$-wave symmetry (giving rise to an effective $p_x$-wave symmetry at the Dirac points) with a strong FVM between the S and N regions. (a) Plot of the length-dependence of the critical current. (b) Plot of the current-phase relationship for $L/\xi=0.1$.}
\label{fig:dwave_doped}
\end{figure}

We proceed to investigate the Josephson current when there is a substantial FVM between the S and N regions. In this case, we obtain from Eq.~(\ref{eq:boundstate_ZES}) that
\begin{align}
I_J&= \frac{e\Delta_0}{\hbar} \sum_{n=0}^N t_n \sqrt{\zeta_n} \sin(\Delta\phi/2)\text{sgn}\{\cos(\Delta\phi/2)\},\notag\\
\end{align}
where $t_n$ models the angular dependence of the gap. For the most interesting $p_x$-wave case, we choose $t_n=\cos(\frac{n\pi}{2N})$. The resulting critical current is shown in Fig.~\ref{fig:dwave_doped}(a), using the parameters $\mu_S/\Delta_0=150$ and $W/\xi=30$. As seen, the current decays like $L^{-1}$ in the regime $\mu_N/\Delta_0 \sim 1-10$ while oscillations appear upon increasing $\mu_N$ further, just as in the $s$-wave case. In agreement with Fig.~\ref{fig:dwave_undoped}(a), it is seen that the oscillation-amplitude is smaller than in the $s$-wave case. The current-phase relationship in the case of a strong FVM for the $d$-wave symmetries is shown in Fig. \ref{fig:dwave_doped}(b).

\subsection{Numerical results}\label{sec:results_numerical}
In this section we will report on the self-consistent numerical results for SNS graphene junctions with both conventional \swave\ and unconventional SB pairing contacts. Our main focus is to complement the analytical work and point to situations where a self-consistent approach is a must in order to capture the correct behavior. We are also able to extract additional data not available through an analytical approach, such as the proximity effect and LDOS, and we will start with reporting these below. This will provide the necessary background to interpret the Josephson current and how it in some cases differs significantly from the analytical result.
\par
As shown above through analytical work, the extended \swave\ solution for SB pairing contacts manifests itself as an on-site pairing gap at non-zero doping, which is always the case in the S regions. We therefore choose to not focus on this solution and only study the two distinct \dwave\ solutions in Eq.~(\ref{eq:sym}) and compare these with on-site \swave\ contacts. More detailed results for the on-site \swave\ solution can be found in earlier work by some of the authors.\cite{black-schaffer_prb_08}
\par
\begin{figure}[h]
\centering
\includegraphics[scale = 0.7]{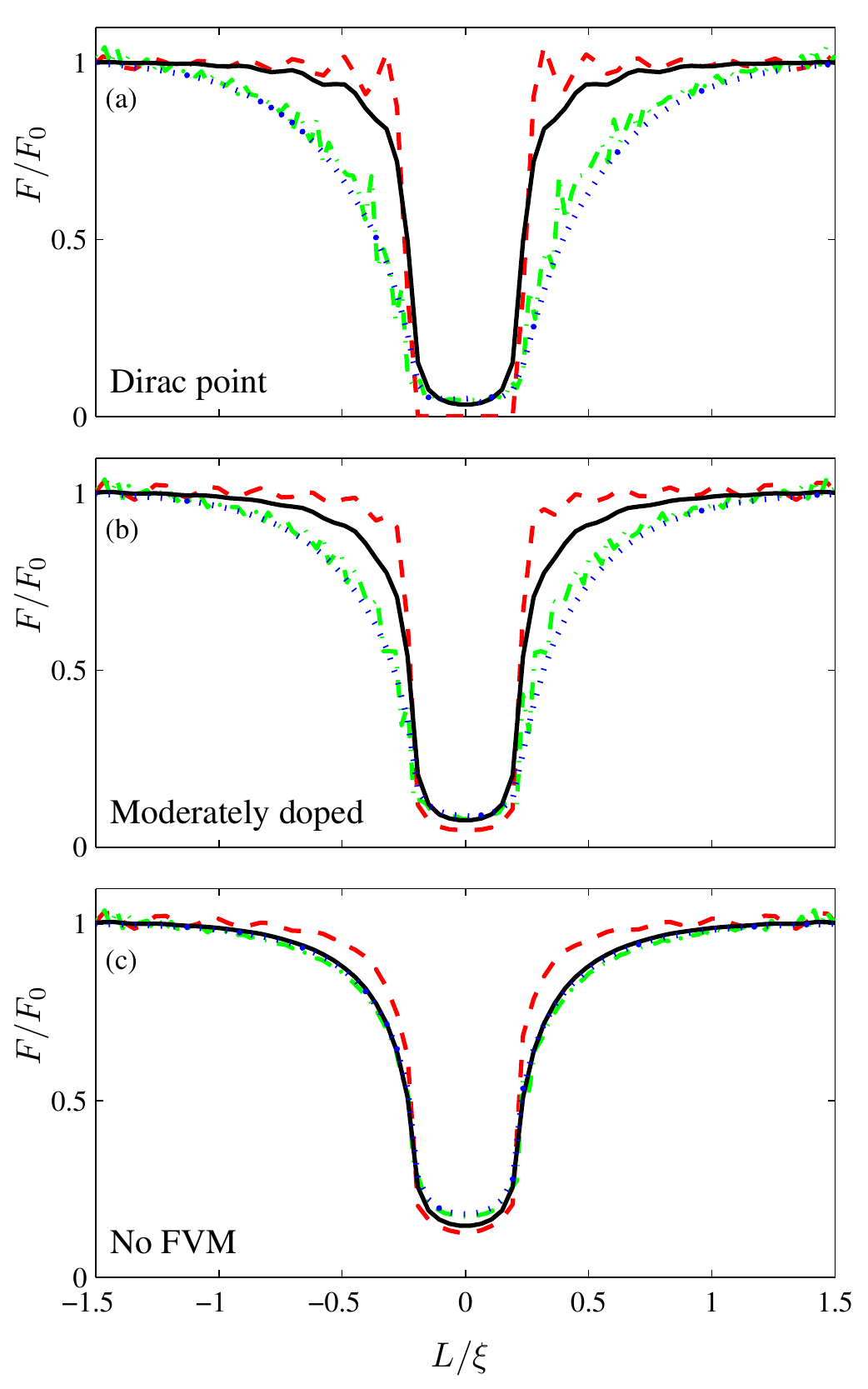}
\caption{(Color online) Proximity effect in terms of normalized pair amplitudes when $\mu$(N) = 0 eV (a), $\mu$(N) = 0.7 eV (b), and $\mu$(N) = 1.5 eV (no FVM) (c). Conventional \swave\ contacts (solid black), \daa\ contacts with zigzag interfaces (dashed red), \dbb\ contacts with zigzag interfaces (dotted blue), and \daa\ contacts with armchair interfaces (dash-dotted green). The width of the N region is $L = 0.42\xi$ and the interfaces are  
marked with vertical black lines.}
\label{fig:proxeff}
\end{figure}

Figure~\ref{fig:proxeff} shows the proximity effect in terms of the normalized pair amplitude for several superconducting symmetries and interface combinations and at multiple doping levels in the N region. 
Solid black lines are the results for on-site \swave\ pairing which show a pronounced depletion of pairs on the S side of the interfaces and the accompanied leakage of pairs into the N regions. This depletion/leakage is stronger at higher doping levels. 
The red dashed curve is the results for the zigzag interface with \daa\ symmetry contacts, i.e.~an effective $p_y$-wave symmetry at low energies, and it displays a much weaker proximity effect compared to the \swave\ contacts. In fact, at zero doping in N there is only a very small, and interestingly, completely flat, non-distance dependent, superconducting pair amplitude inside the junction. The small oscillatory pattern present, especially at lower doping, for this junction can be attributed to charge fluctuations due to the FVM at the interface and is seen also for the \swave\ solution,\cite{black-schaffer_prb_08} though it is less pronounced there.
The green dash-dotted and blue dotted lines are the results for the \daa\ pairing on the armchair interface and the \dbb\ contacts on the zigzag interface, respectively. Both of these have an effective $p_x$-wave symmetry at low energies and should therefore to a good approximation be the same, as is also seen here. We will thus from hereon only report results for the armchair \daa\ pairing. The depletion of pair amplitude on the S side of the interface is notably larger for these contacts compared to \swave\ contacts, but the induced pair amplitude in N is nevertheless not enlarged. Colloquially speaking this means that these contacts have in total lost more pairs than junctions with conventional contacts. 
This effect can be attributed to the formation of quasiparticle ZES at the SN interfaces.
\par
\begin{figure}[h]
\centering
\includegraphics[scale = 0.7]{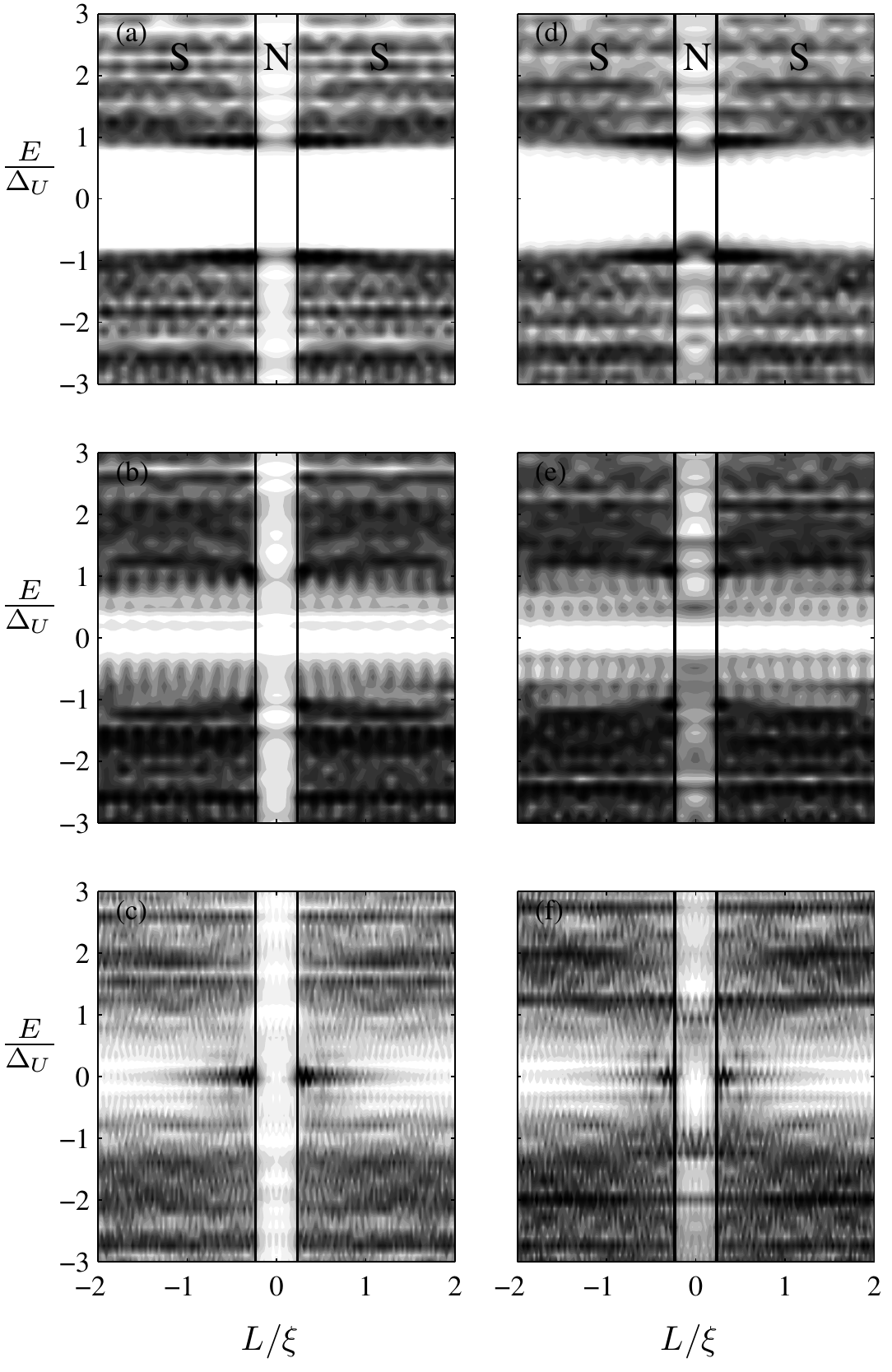}
\caption{LDOS plots for conventional \swave\ contacts (a, d), \daa\ contacts with zigzag interfaces (b, e) and armchair interfaces (c, f). The doping level in the N region is zero (a, b, c) and moderately doped at 0.7 eV (d, e, f). The energy scale has been normalized by the order parameter $\Delta_U$. Black color represent 2 states/eV/unit cell.}
\label{fig:LDOS}
\end{figure}

The presence of ZES for the \dbb\ symmetry on the zigzag interface, or equivalently the \daa\ symmetry on the armchair interface, is the most prominent differences between the two different \dwave\ symmetries in the analytical framework. 
The TB BdG framework allows for a direct access to the LDOS and therefore any ZES formation. Figure~\ref{fig:LDOS} shows the results for $s$-wave contacts (a, d) and $d_{x^2-y^2}$-wave contacts on the zigzag (b, e) and armchair interfaces (c, f). 
For conventional \swave\ contacts we see, as expected, a full gap in the S regions and for short enough junctions this gap persists inside N for both zero (a) and moderately doped (d) N regions. However, the situation is quite different for the \dwave\ contacts. Here, the order parameter in the band structure basis, $C_\vq$ in Eq.~(\ref{eq:Ck}), has nodes on the Fermi surface leading to a familiar V-shaped DOS in the gap. In addition to this feature, we also see pronounced peaks in the LDOS at zero energy at the SN armchair interfaces for \daa\ contacts  as predicted. These ZES are most prominent in the large FVM limit (c) but exists even when the N region is moderately doped and the FVM is smaller (f). At no FVM there is no trace of ZES at the interface. This case corresponds to a diminishing interface barrier $Z$ and even in regular BTK-theory for \dwave\ SN junctions there is then no distinct signatures of the ZES since the transmission probability is unity for $Z = 0$ (see e.g.~Ref.~[\onlinecite{Kashiwaya00}]).
Note that the lighter color throughout the N region in all junctions simply reflect the fact that the DOS is lower here since $\mu_N < \mu_S$.
\par
\begin{figure}[h!tb]
\centering
\includegraphics[scale=0.7]{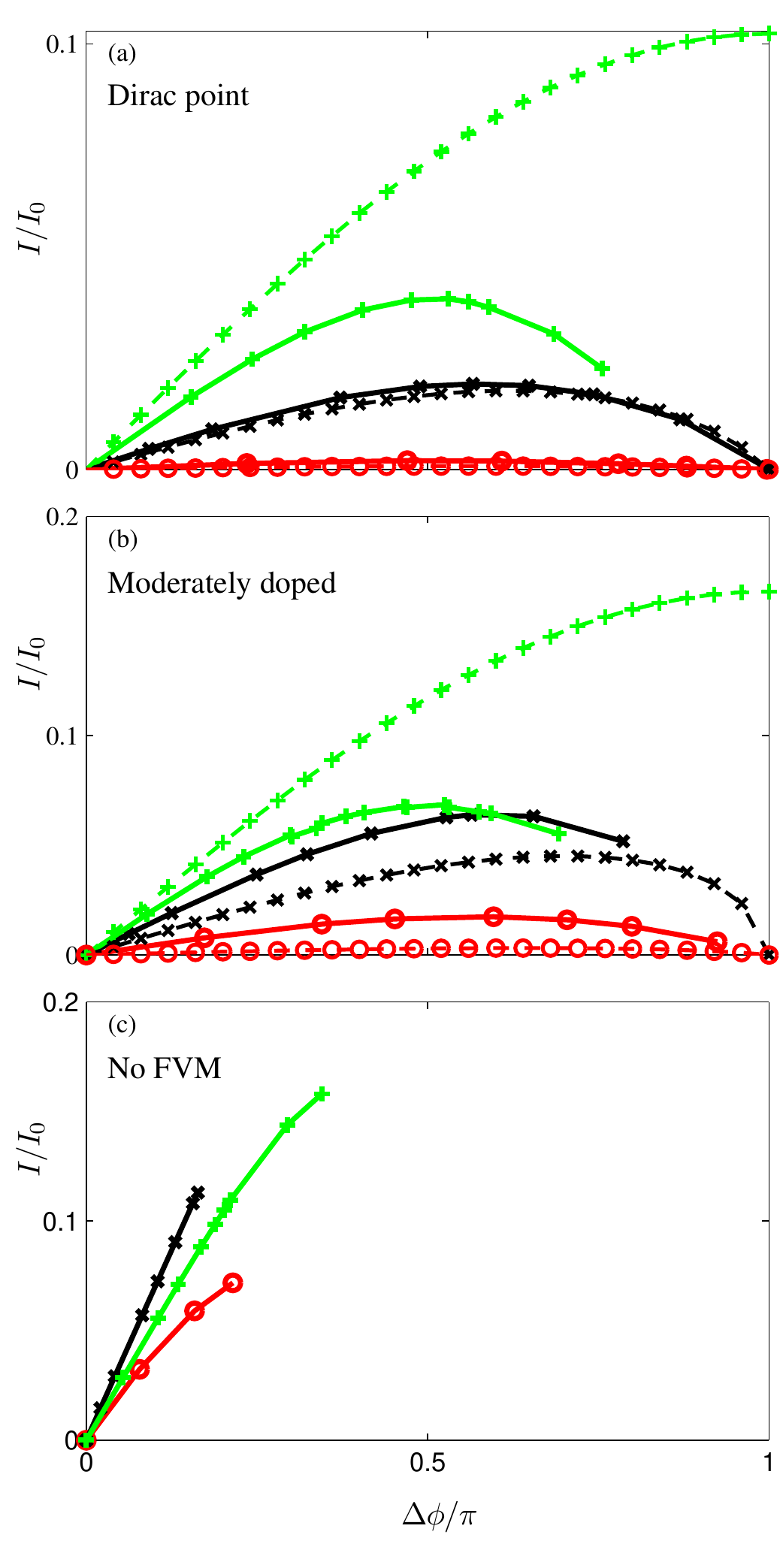}
\caption{(Color online) $I$ vs.~$\Delta \phi$ for $\mu_\text{N}$ = 0 eV (a), 0.7 eV (b), and 1.5 eV (no FVM) (c) for conventional \swave\ contacts (black, $\times$), \daa\ contacts with zigzag interfaces (red, $\circ$) and armchair interfaces (green, $+$). 
The current is given in units of $I_0 = eW/(\hbar \xi)$ when $v_F$ is set to 1.
Analytical results in (a, b) are shown with dashed lines and calculated for the width $W = 30\xi$, which approximates the infinite width of the numerical results.}
\label{fig:Ivsphi}
\end{figure}

In Figs.~\ref{fig:proxeff} and \ref{fig:LDOS} we have seen that different order parameter symmetries in the contacts alter the proximity effect and LDOS significantly and it is therefore expected that there will be a large difference in the Josephson currents as well. Here we will focus on two cases; the dependence of the Josephson current on the phase difference across the junction and on junction length. 
Figure~\ref{fig:Ivsphi} shows the dependence of the Josephson current $I $ on the phase difference $\Delta \phi$ across the junction for different doping levels in N for a junction of length $L=0.42\xi$. 
Here $\Delta \phi$ is the phase difference just across the N region which will always be smaller than the largest phase difference between the two contacts, $\pi$, for any finite Josephson current. This is why the numerical results, especially for larger currents, cannot be extended to large $\Delta \phi$-values.
The numerical results are explicitly calculated for infinitely wide junctions, but by letting $W \rightarrow \infty$ for the analytical results we are able to compare analytical and numerical solutions in the zero and moderately doped cases. In the case of no FVM, the analytical result does not depend on the width and a comparison is much less straightforward and therefore not plotted in Fig.~\ref{fig:Ivsphi}(c).
\par
Let us first focus on the numerical results alone, which are the solid lines in Fig.~\ref{fig:Ivsphi}. For zero doping in N, \daa\ contacts on the armchair interface (green,$+$) have the highest current. This is expected since this is the only configuration with ZES at the SN interfaces and these states will strongly intensify the tunneling current through the junction.\cite{Barash} At least in terms of the Josephson current, the formation of the ZES clearly makes up for the relative loss of pair amplitude seen in Fig.~\ref{fig:proxeff}.
We note that the relative enhancement in the current for \daa\ contacts on armchair interface compared to \swave\ contacts is reduced when the FVM at the interface is reduced. This is also to be expected as the strength of the ZES diminishes with increased doping level in N. In fact, when the ZES disappear at no FVM, the Josephson current is lower than for the similarly strong \swave\ pairing contacts.
Of all the symmetries investigated, \daa\ pairing on the zigzag interface (red, $\circ$) has the lowest current in all three doping regimes. This is consistent with its smallest proximity effect as seen in Fig.~\ref{fig:proxeff}. Finally, also note that the current for all three different pairings increases with increasing doping level in N.
\par
Let us now comment on the explicit $\Delta \phi$ dependence on the current and also make a comparison with the analytical results. The \swave\ contacts show a distinct non-sinusoidal dependence with the numerical results closely tracking the analytical results at low doping levels. At moderately doping the analytical results peak at a higher $\Delta \phi$-value than the self-consistent numerical results, around $0.72\pi$ compared to $0.56\pi$. However, the discrepancy is only moderate, both in terms of critical $\Delta \phi$ and $I$, as already reported in Ref.~[\onlinecite{black-schaffer_prb_08}].
For the \dwave\ contacts the situation is, however, quite different. Especially big is the discrepancy for the case with ZES at the interfaces. Here, the analytical results show a distinct $\text{sgn}\{\cos (\Delta \phi/2)\}\sin (\Delta \phi/2)$ behavior, which is a direct consequence of the presence of the ZES, and a very large critical current for all doping levels. The self-consistent results instead display an almost $\sin (\Delta \phi)$-like curve with a significantly lower critical current. We believe this large difference can be attributed to the importance  
and strength of the ZES. While the ZES are present even in the self-consistent solution and there causes an enhancement of the Josephson current, its importance seem to be significantly reduced within a self-consistent approach. This would explain both the smaller amplitude and the more traditional $\sin (\Delta \phi)$-curve. Since the effect of the ZES is decreased with increasing doping level the discrepancy between the analytical solution and the self-consistent work slightly decreases in terms of maximum current, but the different $\Delta \phi$-dependence will remain as long as the ZES are non-zero in either approach.
It might be worth mentioning here that ZES have been found to be very sensitive to interface properties. Similar numerical solutions schemes as applied here for \dwave\ contacts on the square lattice have found that the ZES can quickly diminish in the presence of random interface potentials, simulating disorder,\cite{asano01}  or for non-straight interfaces destructive interference between different lattice sites can completely kill the ZES\cite{tanuma98}. 
It is therefore very likely that the reduced importance of the ZES in our numerical solution stems from the fact that even a completely smooth, disorder free armchair interface as present in our calculations, is not absolutely flat in the continuum-sense.

However, also the \daa\ symmetry on the zigzag interface shows pronounced differences between the analytical and self-consistent solutions. Here, the phase dependence is similar, with the critical phase difference slightly increasing with doping level in N, but the current for these junctions is significantly enhanced in the self-consistent solution. At the Dirac point there is a $\sim 60\%$ increase in the critical current compared to the analytical solution. This enhancement increases with doping level in N, being $\sim 80\%$ at 0.7 eV doping. It is interesting that we see this enhancement in the current for the self-consistent solution despite the proximity effect causing less leakage into N for these contacts compared to \swave\ pairing. That also means less depletion of the order parameter in the S region of the interface which will give rise to a stronger tunneling and, apparently, this effectively even overcompensates for the relative lack of pair amplitude in N. 
\par
The above shows that for unconventional contacts a self-consistent approach is necessary in order to accurately determine the Josephson current, both in terms of phase dependence and critical current. We have identified at least two causes for this. First, at SN interfaces where ZES are present, a self-consistent calculation is necessary in order to properly evaluate the importance of the ZES to the overall transmission of the junction. Second, the proximity effect itself can be considerably different for  
unconventional contacts as compared to \swave\ contacts. This gives  
rise to the main source of discrepancy in the results between the  
analytical and self-consistent approach in the case of the \daa\  
symmetry on the zigzag interface.
\par
\begin{figure}[h]
\centering
\includegraphics[scale=0.85]{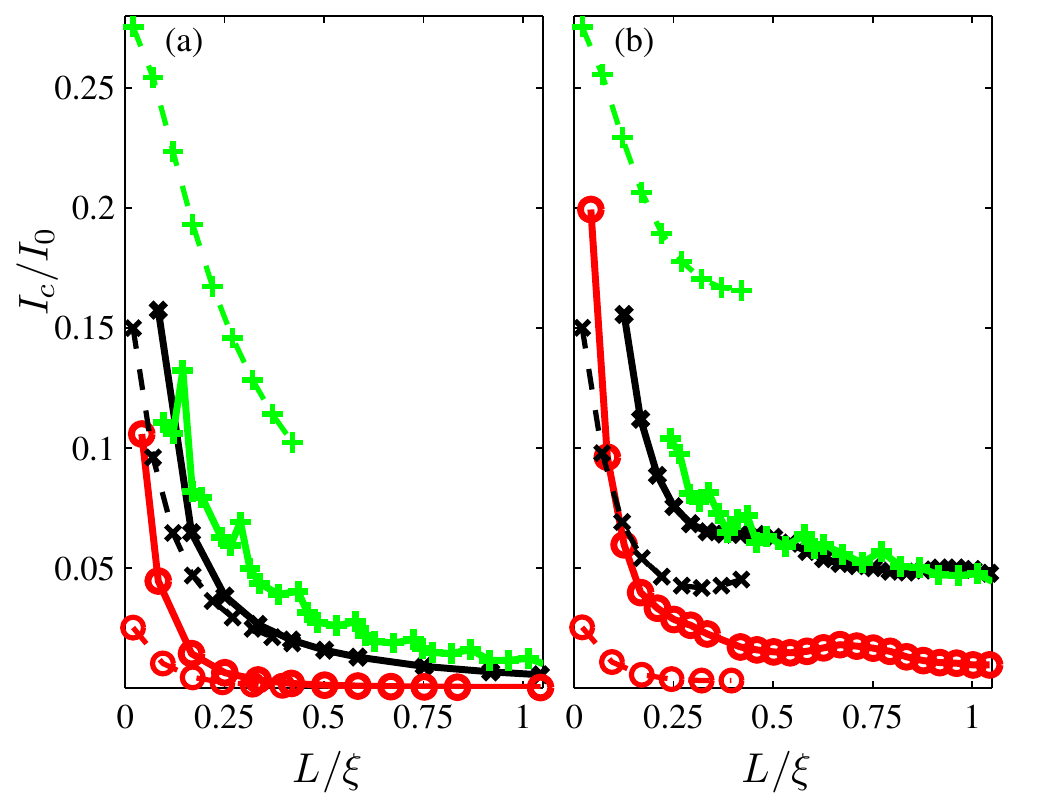}
\caption{(Color online) $I_c$ vs.~$L$ for $\mu_\text{N}$ = 0~eV (a) and 0.7~eV (b) for conventional \swave\ contacts (black, $\times$), \daa\ contacts with zigzag interfaces (red, $\circ$) and armchair interfaces (green, $+$). The current is given in units of $I_0 = eW/(\hbar \xi)$ when $v_F$ is set to 1.
Analytical results are shown with dashed lines and calculated for the width $W = 30\xi$, which approximates the infinite width of the numerical results.}
\label{fig:IvsL}
\end{figure}

Next, we report on the junction length dependence of the Josephson current. The analytical approach used here is only reliable in the short junction regime where the current is mainly carried by the Andreev bound states. For the self-consistent TB BdG approach such a limitation does not exist and we can therefore extend the results to much longer junctions. 
On the other hand, limited computational time requires relatively short superconducting coherence lengths and with the parameters used here the shortest junctions we can study have $L=0.04\xi$. Also, for these short junctions there will be a substantial current across the junction and, as a result, the available phase difference $\Delta \phi$ across the junction is drastically reduced making it hard to reach $\Delta \phi_c$ for very short junctions. This means that we effectively do not have any results for junctions shorter than $\sim 0.1\xi$ for $\mu = 0$~eV and $\sim 0.25\xi$ for $\mu=0.7$~eV. For relatively large junction lengths the analytical results unfortunately start to become questionable and we can therefore not make a comparison between the analytical and the self-consistent results over any extended length scale.
Figure \ref{fig:IvsL} shows the critical current $I_c$ vs.~length of junction $L$ for all three cases of symmetries of the pairing at doping levels $0$~eV (a) and 0.7~eV (b) in N. 
At zero doping in N (a) \swave\ contacts show a relatively good agreement between the analytical and self-consistent results, although the length dependence is somewhat stronger in the self-consistent case.\cite{black-schaffer_prb_08} A simple least square fit to the functional form $I_c = CL^{-\beta}$ gives $\beta = 1.3$ for the self-consistent curve but only 0.8 for the analytical results.
The \daa\ contacts on the zigzag interface show a similar increased length dependence for short junctions with $\beta = 2$ for the self-consistent results and 1.3 for the analytical counterpart. 
The \daa\ contacts on the armchair interface have the weakest $L$-dependence with only $\beta = 1.1$ for the self-consistent results. The analytical results in this case give a bad fit to a power-law dependence as the critical current levels off for short junctions.
There are no noticeable oscillations in the critical current as a function of junction length in the undoped regime for either of \swave\ or the \daa\ contacts with the zigzag interfaces for either solution method. However, \daa\ contacts with the armchair interface exhibit peaks in the critical current at every 6 unit cells or $0.14\xi$ in the self-consistent approach. These are most pronounced for short junctions, but also exist for the longest junctions we have studied. Since the analytical solution does not capture these oscillations, we  
attribute the peaks to resonance transmission due to changes in the  
relative strength or interactions of the ZES.
\par
At moderate doping and thus a smaller FVM (b) we see a fairly large discrepancy in the critical value of the current between the analytical and the self-consistent solutions for all three contacts, as also seen in Fig.~\ref{fig:Ivsphi}. The length dependence is however very similar in the two solutions for both \swave\ and \daa\ contacts on the zigzag inteface. Here $\beta = 0.5$ and 0.85, respectively. For the junction with ZES, the length dependence is harder to estimate because of the lack of short junction self-consistent data but the results approximately vary from $\beta = 0.2$ to 0.5 when self-consistency is included.
At moderate doping both the analytical and the self-consistent data show long wave length oscillations for especially the $s$- and $d_{x^2-y^2}$-wave contacts with the zigzag interfaces. While it is not straightforward to compare frequencies and amplitudes of the analytical and self-consistent approaches there seem to be an overall agreement.
For \daa\ contacts on the armchair interface, we again find short wave lengths resonance transmission peaks in the self-consistent solution, but we can in this case not easily identify any distinct frequency. 
\par
\begin{figure}[hbt]
\centering
\includegraphics[scale=0.7]{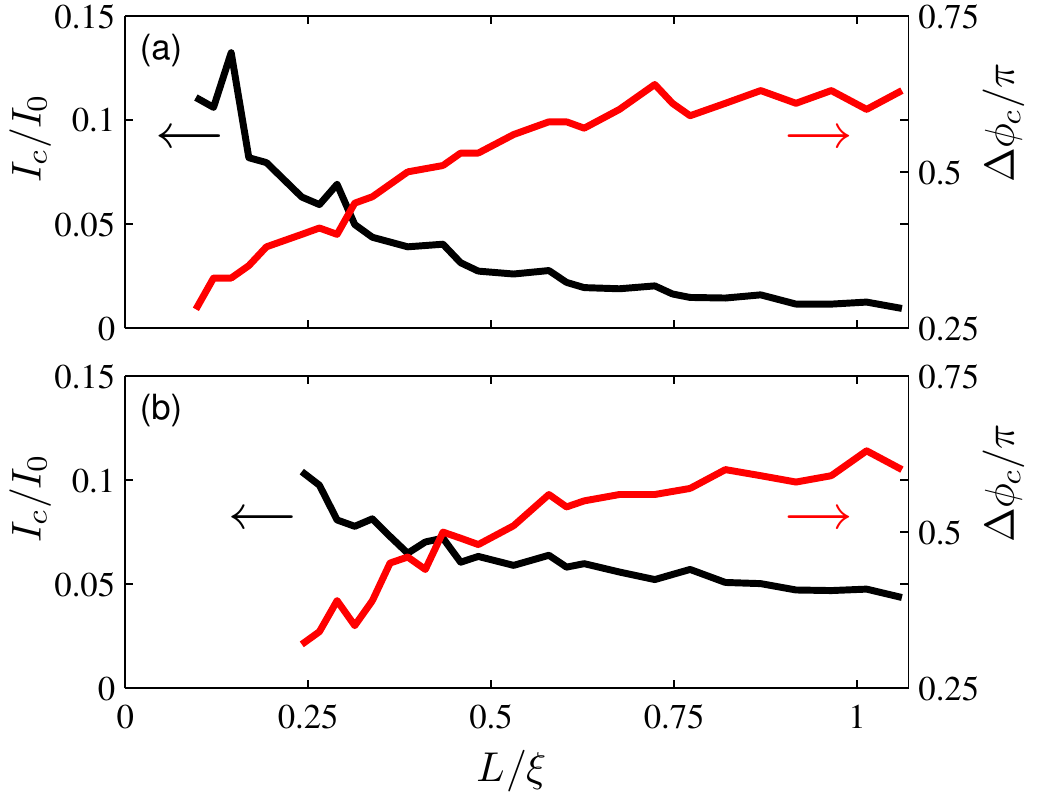}
\caption{(Color online) $I_c$ (black, left axis) and $\Delta \phi_c$ (red,  
right axis) vs.~$L$ for $\mu_\text{N}$ = 0~eV (a) and 0.7~eV (b) for  
the \daa\ solution on the armchair interface.}
\label{fig:Ic_phic}
\end{figure}

Not only do the $I$ vs.~$\Delta \phi$ and $I_c$ vs.~$L$ curves differ most for the \daa\ contacts on the armchair interface, but Fig.~\ref{fig:Ic_phic} shows that there is a strong junction length dependence on the critical phase difference $\Delta \phi_c$ where the maximum current is reached. 
As seen in Figs.~\ref{fig:dwave_undoped}-\ref{fig:dwave_doped}, $\Delta \phi_c = \pi$ for all junction lengths in the analytical solution in this case, whereas the self-consistent solution shows $\Delta \phi_c$ increasing from around $0.3\pi$ in the shortest junctions we could study to above $0.6\pi$ in the longest junctions. The spread in $\Delta \phi_c$ could be even larger, but the curves seem to level out at large junction lengths. Interestingly, this does not only mean that the maximum current is not reach at a $\pi$ phase difference, but also that the phase difference is advanced compared to the conventional $\sin(\Delta \phi)$ curve for short junctions.
It might be worth noting that there is a fair amount of fluctuations in the data in Fig.~\ref{fig:Ic_phic}. Part of this is due to the way we measure the Josephson current. We apply a fixed phase difference across the whole SNS structure, thus technically we are only able to give an interval for the critical current and phase difference. However, the peaks in critical current seen in undoped junctions for every 6 unit cell are not a numerical artifact, and as seen, they are also not correlated with the wiggles in the critical phase difference. We admit that it is somewhat harder to draw the same conclusion in the moderately doped case.
There is no similar junction length dependence for \swave\ or \daa\ pairing on the zigzag interface, but here $\Delta \phi_c$ is fixed over the whole range of $L$ we investigate.

\section{Conclusions}
\label{sec:summary}
In summary, we have studied the Josephson current in a graphene superconductor/normal/superconductor (SNS) junction. Superconductivity is induced in graphene by means of the proximity effect from a superconducting host material. In particular, we assume that superconducting contacts are deposited on top of the graphene sheet as realized experimentally.\cite{heersche_nature_06} 
Whereas previous works on the Josephson current in graphene have mainly treated only an isotropic $s$-wave symmetry for the superconducting order parameter,\cite{beenakker_prl_06, titov_prb_06, bhattacharjee_prl_06, moghaddam_prb_06, linder_prl_07, black-schaffer_prb_08, bhattacharjee_prb_07, maiti_prb_07, yokoyama_prb_08, rainis_prb_09, liang_prl_08}  we here also present an analysis for anisotropic pairing that arises due to nearest-neighbor interactions. This latter pairing can either have an extended \swave\ symmetry or belong to any linear combination of $d_{xy}$ and $d_{x^2-y^2}$. 
\par
We show that a junction with extended \swave\ symmetry, which displays an effective $p_x+\i p_y$-wave symmetry near the Dirac points, is equivalent to a junction with on-site, isotropic \swave\ pairing. While \swave\ pairing has been studied before, we here report on newly found oscillations in the critical current as function of junction length in both junctions with no Fermi vector mis-match (FVM) at the SN interfaces and in heavily doped junctions with a strong FVM.
\par
For the case of \dwave\ superconducting contacts we limit our investigation to considering only  $d_{x^2-y^2}$-wave pairing on the zigzag and armchair interfaces. Since all pairings are induced from on-top deposited contacts, different symmetry choices simply correspond to different orientations of the contacts relative to the graphene sheet.
These chosen symmetries give rise to effective $p_x$- and $p_y$-wave pairing, respectively, at low energies. Therefore \daa\ pairing on the armchair interface is at low energies equivalent to \dbb\ pairing on the zigzag interface and so on, making our study quite general.
In an experimental setup any chiral interface on the graphene could be realized. However, we believe that this situation can at least qualitatively be predicted from our results and will mainly depend on the presence of zero energy states (ZES) at the interfaces. These states appear if the order parameter changes sign when the angle of incidence for the quasiparticles on the SN interface is changed from $\theta$ to $\pi-\theta$ as is the case for \daa\ pairing on the armchair interface.
\par
We calculate the Josephson current both analytically and numerically by a self-consistent approach for all the above symmetries. Whereas there is good agreement between the two treatments for the \swave\ superconducting order parameters, there is a pronounced deviation between the two methods for anisotropic pairing, in particular, when ZES are present. These states at zero energy will easily dominate the transport through the junction, and while they are present even in the numerical results, their effect on the Josephson current is strongly suppressed when self-consistency is achieved.
One easily identified source for this deviation is the first order expansion to $p$-wave symmetries around the Dirac points done in the analytical treatment. As seen in Fig.~\ref{fig:gaps}, the \dwave\ order parameters differ from pure $p$-waves at higher energies. Since the contacts are likely to induce a rather heavy doping into the graphene in the S regions one might argue that this effect is not negligible. However, as seen in Fig.~\ref{fig:proxeff}, the proximity effect is remarkably similar for the \dbb\ on zigzag interface and the \daa\ on the armchair interface, thus pointing to the fact that this is not a major source of deviation between the analytical and the self-consistent treatment. It is instead the self-consistency for the proximity effect inside the junction that is the crucial component. Therefore, a numerical, self-consistent calculation is required in order to properly address the transport properties of graphene when the superconducting pairing is anisotropic in $\vk$-space, especially when zero-energy states are present at the interface.
\section*{Acknowledgments}
J.L. is grateful to M. Titov for a helpful e-mail correspondance. H. Haugen is thanked for useful discussions. A.M.B.-S. acknowledges support from the Department of Applied Physics and the School of Humanities and Sciences, Stanford University.
J.L. and A.S. were supported by the Research Council of Norway, 
Grants No. 158518/431 and No. 158547/431 (NANOMAT), and Grant No. 167498/V30 (STORFORSK). 
T. Y. acknowledges support by JSPS. 


\end{document}